\documentclass[english,journal]{IEEEtran}
\usepackage[T1]{fontenc}
\usepackage[latin9]{inputenc}
\usepackage{float}
\usepackage{amsthm}
\usepackage{amsmath}
\usepackage{amssymb}
\usepackage{graphicx}

\makeatletter

\floatstyle{ruled}
\newfloat{algorithm}{tbp}{loa}
\providecommand{\algorithmname}{Algorithm}
\floatname{algorithm}{\protect\algorithmname}

\theoremstyle{plain}
\newtheorem{thm}{\protect\theoremname}
\theoremstyle{definition}
\newtheorem{defn}[thm]{\protect\definitionname}
\theoremstyle{definition}
\newtheorem{problem}[thm]{\protect\problemname}

\@ifundefined{showcaptionsetup}{}{%
 \PassOptionsToPackage{caption=false}{subfig}}
\usepackage{subfig}
\makeatother

\usepackage{babel}
\providecommand{\definitionname}{Definition}
\providecommand{\problemname}{Problem}
\providecommand{\theoremname}{Theorem}

\begin{document}

\title{Local Thresholding in General Network Graphs}

\author{Ran Wolff\\
Information Systems Department \\
University of Haifa\\
rwolff@is.haifa.ac.il}

\maketitle
Local thresholding algorithms were first presented more than a decade
ago and have since been applied to a variety of data mining tasks
in peer-to-peer systems, wireless sensor networks, and in grid systems.
One critical assumption made by those algorithms has always been cycle-free
routing. The existence of even one cycle may lead all peers to the
wrong outcome. Outside the lab, unfortunately, cycle freedom is not
easy to achieve. 

This work is the first to lift the requirement of cycle freedom by
presenting a local thresholding algorithm suitable for general network
graphs. The algorithm relies on a new repositioning of the problem
in weighted vector arithmetics, on a new stopping rule, whose proof
does not require that the network be cycle free, and on new methods
for balance correction when the stopping rule fails. 

The new stopping and update rules permit calculation of the very same
functions that were calculable using previous algorithms, which do
assume cycle freedom. The algorithm is implemented on a standard peer-to-peer
simulator and is validated for networks of up to 80,000 peers, organized
in three different topologies representative of major current distributed
systems: the Internet, structured peer-to-peer systems, and wireless
sensor networks.

\section{Introduction}

Recent years have seen a surge in the number, the pervasiveness, and
the capabilities of networked devices, followed by ever greater interest
in efficient algorithms for distributed computation of various functions.
These functions can be the main application of the system (e.g., target
tracking), a necessary preprocessing stage of an application (e.g.,
outliers removal), or a subsystem (e.g., a load-balancing service).
Some of this interest is driven by the performance superiority of
in-network computation. Performance becomes increasingly important
as distributed data becomes abundant (e.g., when processed by apps
running on smartphones), as the balance between computation and communication
costs tilts in favor of the former (e.g., in a peer-to-peer environment),
and when energy conservation becomes a major concern (as in a wireless
sensor network). Additional causes for this interest are more social
than technological: a distributed architecture -- harder to manipulate
and control by any single entity -- is preferred where users distrust
any third party, and where privacy is a concern. This is frequently
the case where there is suspicion of bias (e.g., in shopping recommendations)
or in semi-legitimate applications (e.g., file sharing).

In-network computation algorithms fall into several categories. Those
categories which provide proof of correctness include broadcast and
convergecast based algorithms such as those implemented using MapReduce
\cite{disco} or tailor-made methods \cite{Unnamed-3}, gossip algorithms
\cite{KempeGossip,gossipBoyd,Gossip,gossipSpeedup,GossipVoting,gossipKalman,gossipKalman1,gossipSL,dynamicGossip},
and local thresholding algorithms \cite{MajorityRulej,localRanking,l2,p2pDT,p2pEigen}.
Broadcast and convergecast require global coordination, which can
be costly. When dealing with dynamically changing data and topologies,
as in peer-to-peer and wireless sensor networks, such algorithms are
not attractive. Gossip algorithms require no coordination and are
suitable for dynamic data and topologies. However, the correctness
of gossip algorithms relies on rapid mixing of the inputs, which is,
by definition, communication intensive. 

Local thresholding algorithms are a third category of in-network computation
algorithms. Unlike gossip algorithms, they focus on decision rather
than approximation problems. For instance, a basic local thresholding
algorithm \cite{MajorityRulej} would compute a majority vote where
its counterpart gossip based algorithm computes the average. Research
has shown that many data mining algorithms (e.g., a priori association
rule mining, ID3 for decision tree induction, and approximated versions
of k-mean clustering) can be mapped to large numbers of convex thresholding
decisions.

At the heart of any local thresholding algorithm lies a local stopping
rule: a condition computed by each peer on its data and the messages
it has received and sent. When the condition is violated, the peer
must send out messages in prescribed ways. However, when the condition
holds, the peer does not have to send any further messages because
either every peer currently computes the correct outcome, or there
is a peer in the network whose condition is violated and who is responsible
for correcting the computation outcome of all peers. Since local thresholding
algorithms rely on achieving local balance rather than on mixing the
inputs, they are the most communication thrifty of the three categories.
However, all local thresholding algorithms presented until today require
cycle-free routing, which makes them very difficult to implement in
real distributed systems. 

This work is the first to present a local thresholding algorithm suitable
for general network graphs. The algorithm relies on a new local stopping
rule and on new update rules that restore local balance when the stopping
rule fails. Unlike those used in previous local thresholding algorithms,
the new rules do not rely on cycle freedom for their correctness.
The new stopping and update rules are general and can replace any
of those used in existing local data mining algorithms. Additionally,
because they can handle cycles, algorithms based on the new rules
can handle partial message failure. Of no less importance is our representation
of the problem in a new mathematical framework, which simplifies proofs
and invites further development of local thresholding algorithms.

The rest of this paper is organized as follows: Section \ref{sec:Notation-and-Preliminaries}
describes the notation used and the formal problem definition, as
well as the success metrics. Section \ref{sec:A-local-stopping} provides
the main mathematical contribution -- the new stopping rule. Section
\ref{sec:Balance-correction} complements Section \ref{sec:A-local-stopping}
by describing a general and accurate balance correction method. Section
\ref{sec:Source-selection} combines those two contributions to form
a useful algorithm. Thorough experimentation is described in Section
\ref{sec:Experimental-evaluation}. Section \ref{sec:Related-work}
explains the relation of previous work to this one. We conclude with
some open research questions in Section \ref{sec:Conclusion-and-further}.

\section{\label{sec:Notation-and-Preliminaries}Notations and problem statement}

\subsection{\label{sub:Mathematical-notation}Mathematical notation}

Much of the math in this paper uses weighted averages. To simplify
notation throughout the paper we adapt the notation proposed in \cite{dynamicGossip}
in which $\oplus$ denotes the weighted average of a pair and $\bigoplus$
the weighted average of a set of vectors. The scalar multiplication
of a weighted vector, which affects its weight, is denoted $\odot$.
A formal definition follows.
\begin{defn}
{[}Weighted Vector Space{]} Let $\mathcal{V}$ be a vector space and
$\mathcal{C}$ a corresponding field of scalars. Denote $+$ and $\cdot$
the addition and the scalar multiplication of $\mathcal{V}$. The
weighted vector space $\mathcal{W}$ with addition $\oplus$ and scalar
multiplication $\odot$ is defined as follows:\end{defn}
\begin{itemize}
\item The elements of $\mathcal{W}$ are pairs $\left\langle \overrightarrow{v},c\right\rangle $
such that $\overrightarrow{v}\in\mathcal{V}$ and $c\in\mathcal{C}$.
\item The scalar field of $\mathcal{W}$ is $\mathcal{C}$.
\item The scalar multiplication is defined: 
\[
c_{1}\odot\left\langle \overrightarrow{v},c_{2}\right\rangle \doteq\left\langle \overrightarrow{v},c_{1}\cdot c_{2}\right\rangle .
\]

\item The addition is defined: 
\[
\left\langle \overrightarrow{v_{1}},c_{1}\right\rangle \oplus\left\langle \overrightarrow{v_{2}},c_{2}\right\rangle \doteq\left\langle \frac{c_{1}}{c_{1}+c_{2}}\cdot\overrightarrow{v_{1}}+\frac{c_{2}}{c_{1}+c_{2}}\cdot\overrightarrow{v_{2}},c_{1}+c_{2}\right\rangle .
\]

\end{itemize}
To be precise, $\frac{c_{1}}{c_{1}+c_{2}}$ denotes the multiplication
of $c_{1}$ in the scalar inverse of $c_{1}+c_{2}$. The obvious example
for a weighted vector space is one in which $\mathcal{V}$ is $\mathbb{R}^{d}$
and $\mathcal{C}$ is $\mathbb{R}$. However, the definition of the
weighted vector space is general enough to include many types of $\mathcal{V}$
and $\mathcal{C}$. Of special interest is that for a specific space
of random vectors $\mathcal{V},$ the corresponding $\mathcal{C}$
can be the space of covariance matrices. This has many applications
in data mining and machine learning, including z-score normalization.

It is easy to validate that the weighted vector space $\mathcal{W}$
with the operations $\odot$ and $\oplus$ is a vector space and that
any $X_{0}$ whose weight is the zero element of $\mathcal{C}$ is
an identity element of this space. Also, the triangle inequality with
respect to the $L2$ norm $\left\Vert \cdot\right\Vert $ holds for
the vector part of the weighted vector. 

For brevity, we make four additional notations: We will refer to the
vector part of any $X\in\mathcal{W}$ as $\overrightarrow{X}$ and
to its scalar part as $\left|X\right|$. Additionally, the additive
inverse operator is denoted by $\ominus$ where $X\ominus Y=Z$ if
and only if $X=Y\oplus Z$%
\footnote{The $\ominus$ operator requires careful use since $X\ominus Y$ is
undefined when $\left|X\right|=\left|Y\right|$.%
}. Finally, for a set $\chi=\left\{ X_{1},X_{2},\dots,X_{n}\right\} $
with $X_{i}\in\mathcal{W},$ the additive iteration $X_{1}\oplus X_{2}\oplus\dots\oplus X_{n}$
is shorthanded to $\bigoplus_{X_{i}\in\chi}X_{i}$.

\subsection{Problem statement}

Let $P=\left\{ p_{1},\dots,p_{n}\right\} $ be a set of peers with
inputs $X=\left\{ X_{1,1},\dots,X_{n,n}\right\} $, respectively.
Let $N_{i}\subseteq P$ be the set of peers connected to $p_{i}$.
The average input is $\bigoplus_{p_{i}\in P}X_{i,i}$ or just $\bigoplus X$
for short. Peers communicate by sending messages, where each message
consists of a single weighted vector. The latest message sent by $p_{i}$
to a neighbor $p_{j}\in N_{i}$ is denoted $X_{i,j}$. Unlike previous
art, we do not assume any structure on the communication graph aside
from connectedness%
\footnote{If the network is disconnected, then any connected component carries
an independent computation.%
}.

Throughout this paper it is assumed that the set of peers, $P$, their
inputs, $X$, and the connectivity of each peer, $N_{i}$, vary over
time. Link failures are modeled as changes in the neighbor sets of
the peers at both ends of the link and peer failure as failure of
all its links. The method of failure detection is not specified and
it is sufficient that failures are eventually detected (i.e., a heartbeat
mechanism is sufficient.) This paper assumes symmetric communication,
i.e., that $p_{j}\in N_{i}\Leftrightarrow p_{i}\in N_{j}$. Our proofs
of correctness also assume communication is ordered and reliable.
However, we show (in Section \ref{sec:Source-selection}) how sufficient
ordering can be enforced. We further show experimentally that limited
random message dropping does not affect correctness in any serious
way.

The objective of all the peers is to compute a function $f\left(\overrightarrow{\bigoplus X}\right)$.
Let $f_{i}$ be a function which determines the output of peer $p_{i}$.
An algorithm provides \emph{eventual correctness} if, whenever changes
cease for a long enough period, it guarantees all $f_{i}$ will converge
to $f\left(\overrightarrow{\bigoplus X}\right)$ as computed on the
current set of inputs. Often, however, changes never cease for a long
enough period, and intermittent accuracy -- the percent of peers which
compute the correct outcome -- is more important than convergence.
An algorithm is local if the resources every peer requires in order
to arrive at a prescribed level of intermediate accuracy tend to a
constant when the number of peers tends to infinity.

Under the said assumptions we denote the agreement of $p_{i}$ and
its neighbor $p_{j}\in N_{i}$ by $A_{i,j}=X_{i,j}\oplus X_{j,i}$.
We note that unless messages are still traveling between $p_{i}$
and $p_{j}$, $A_{i,j}=A_{j,i}$. The state of $p_{i}$ is denoted
$S_{i}=X_{i,i}\oplus\left(\bigoplus_{p_{j}\in N_{i}}\left(X_{j,i}\ominus X_{i,j}\right)\right)$.
Finally, in this paper we are interested in a specific family of problems:
\begin{problem}
Let $\mathcal{R}=\left\{ R_{1},R_{2},\dots\right\} $ be a (possibly
infinite) set of non-overlapping convex regions in $\mathbb{R}^{d}$
and let $f\left(\overrightarrow{X}\right)=\begin{cases}
R & \overrightarrow{X}\in R:R\in\mathcal{R}\\
nil & otherwise
\end{cases}$. The problem is to compute, at every $p_{i}$, 
\begin{eqnarray*}
f_{i}\left(X_{i,i},\left\{ X_{i,j},X_{j,i}:p_{j}\in N_{i}\right\} \right) & = & f\left(\overrightarrow{\bigoplus X}\right).
\end{eqnarray*}

\end{problem}
As stated earlier, many data mining problems can be reduced to this
generic problem. The solution provided in this paper is to compute
at every peer a status $S_{i}$ which guarantees 
\begin{eqnarray*}
f_{i}\left(X_{i,i},\left\{ X_{i,j},X_{j,i}:p_{j}\in N_{i}\right\} \right) & = & f\left(\overrightarrow{S_{i}}\right)=f\left(\overrightarrow{\bigoplus X}\right).
\end{eqnarray*}

\section{\label{sec:A-local-stopping}A local stopping rule for general network
graphs}

In this section we prove the main result of the paper: a new stopping
rule which does not require that the network be cycle free. We first
show that throughout the workings of the algorithm, the average of
the inputs is reflected in the states the different peers maintain.
This provides that the global input is preserved in the states regardless
of how it is distributed by the algorithm.
\begin{thm}
\label{lem:The-global-sum}{[}Mass Conservation (\cite{KempeGossip}
Proposition 2.2){]} The average of the states of all peers is equal
to the average of the inputs of all peers; i.e., $\bigoplus_{p_{i}\in P}S_{i}=\bigoplus X$.\end{thm}
\begin{IEEEproof}
\begin{eqnarray*}
\bigoplus_{p_{i}\in P}S_{i} & = & \bigoplus_{p_{i}\in P}\left(X_{i,i}\oplus\left(\bigoplus_{p_{j}\in N_{i}}\left(X_{j,i}\ominus X_{i,j}\right)\right)\right)=
\end{eqnarray*}
\[
\left(\bigoplus_{p_{i}\in P}X_{i,i}\right)\oplus\left(\bigoplus_{p_{i}\in P}\left(\bigoplus_{p_{j}\in N_{i}}X_{j,i}\right)\right)\ominus\left(\bigoplus_{p_{i}\in P}\left(\bigoplus_{p_{j}\in N_{i}}X_{i,j}\right)\right).
\]
Since every $X_{i,j}$ appears twice, once preceded by $\oplus$ and
once by $\ominus$, we have that 
\begin{eqnarray*}
\left|\left(\bigoplus_{p_{i}\in P}\left(\bigoplus_{p_{j}\in N_{i}}X_{j,i}\right)\right)\ominus\left(\bigoplus_{p_{i}\in P}\left(\bigoplus_{p_{j}\in N_{i}}X_{i,j}\right)\right)\right| & = & 0.
\end{eqnarray*}
 In other words, it is an identity element. It follows that for some
$X_{0}$ with $\left|X_{0}\right|=0$
\begin{eqnarray*}
\bigoplus_{p_{i}\in P}S_{i} & = & \left(\bigoplus_{p_{i}\in P}X_{i,i}\right)\oplus X_{0.}\\
 & = & \bigoplus X
\end{eqnarray*}

\end{IEEEproof}
The center of every local thresholding algorithm is a local stopping
rule: a condition, calculable independently by every peer, subject
to which that peer can stop sending messages to its neighbors. The
stopping rule used in this paper is the following:
\begin{defn}
\label{[Local-stopping-rule]}{[}Local-stopping-rule{]} A peer $p_{i}$
can stop sending messages in the context of a convex region $R\subset\mathbb{R}^{d}$
if the following two conditions both hold with respect to every $p_{j}\in N_{i}$:
\begin{itemize}
\item Either $\left|A_{i,j}\right|=0$ or $\overrightarrow{A_{i,j}}\in R$
.
\item Either $\left|S_{i}\ominus A_{i,j}\right|=0$ or $\overrightarrow{S_{i}\ominus A_{i,j}}\in R.$
\end{itemize}
\end{defn}
When the conditions defined above are satisfied, $p_{i}$ sends no
message unless the rule is once more violated as a result of an incoming
message or a change to $p_{i}$'s input. The situation in which the
stopping rule is satisfied for all peers and in which no further messages
traverse the network is denoted a stopping state. Since we assume
the data does change, it is not a termination state. Nevertheless,
correctness is required and is proven for stopping states. Next, we
prove the first part of the claim for correctness: that in every stopping
state the status vectors $\overrightarrow{S_{i}}$ of all peers reside
in the same region.
\begin{thm}
\label{lem:[Consensus]}{[}Consensus{]} Let $\mathcal{R}=\left\{ R_{1},R_{2}\dots\right\} $
be a set of non-overlapping regions in $\mathbb{R}^{d}.$ In every
stopping state there is some $R\in\mathcal{R}$ such that for all
$p_{i},$ $\overrightarrow{S_{i}}\in R$.\end{thm}
\begin{IEEEproof}
Since no messages traverse the network and assuming reliable communication,
$\overrightarrow{A_{i,j}}=\overrightarrow{A_{j,i}}$ in every stopping
state. Since the regions are non-overlapping, if $\overrightarrow{S_{i}}$
and $\overrightarrow{S_{j}}$ are in two different regions then they
cannot both be in the same region with $\overrightarrow{A_{i,j}}$.
It follows that $\overrightarrow{S_{i}}$, $\overrightarrow{S_{j}}$,
$\overrightarrow{A_{i,j}}$, and $\overrightarrow{A_{j,i}}$ are all
in the same region. Since the graph is connected, this equivalence
transitions to all peers. 
\end{IEEEproof}
Next we prove our main theorem, which is that under the local stopping
rule in Definition \ref{[Local-stopping-rule]}, which makes no assumptions
on the network graph properties, the status vector $\overrightarrow{S_{i}}$
of all peers resides in the same region $R$ which contains $\overrightarrow{\bigoplus X}$.
Relying on Theorem \ref{lem:[Consensus]}, the Theorem below concerns
just one region -- the one agreed on by all peers in the stopping
state.
\begin{thm}
\label{thm:main}{[}Local Stopping Rule{]} If in a stopping state
the status vectors $\overrightarrow{S_{i}}$ of all $p_{i}\in P$
reside in some convex $R\subseteq\mathbb{R}^{d}$, then $\overrightarrow{\bigoplus X}\in R$
as well.
\end{thm}
The proof of Theorem \ref{thm:main} is by describing a possible sequence
of messages which may follow in every stopping state. After each message
is sent and received, the state of the systems remains a stopping
state. We stress that although messages in our proof traverse a tree,
this is merely a proving methodology and not an assumption on the
structure of the network.
\begin{IEEEproof}
First, note that if both $\overrightarrow{A_{i,j}}$ and $\overrightarrow{S_{i}\ominus A_{i,j}}$
are in some $R\in\mathcal{R}$ and since all $R\in\mathcal{R}$ are
convex, $\overrightarrow{S_{i}}=\overrightarrow{S_{i}\ominus A_{i,j}\oplus A_{i,j}}$
must be in $R$ as well. It thus follows from Theorem \ref{lem:[Consensus]}
that all $p_{i}$ agree on the same $R$. Let $T\left(P,E\right)$
be a spanning tree over the network graph rooted at an arbitrary peer.
Consider a convergecast process in which, starting at the leaves,
every peer $p_{i}$ waits until it receives a message from all of
its descendants and then sends a message to its parent $p_{j}$. The
content of the message is selected such that the new value of $A_{i,j}$
becomes $A_{i,j}\oplus S_{i}$.

First, consider a leaf $p_{i}$. The leaf waits for no incoming messages
and sends a message to its parent $p_{j}$. To set the new $A'_{i,j}$
to $S_{i}$, the content of the message should be $S_{i}\oplus X_{i,j}$.
This way, $A'_{i,j}=S_{i}\oplus X_{i,j}\oplus X_{j,i}=S_{i}\oplus A_{i,j}$.
Sending the message adds $S_{i}$ to $X_{i,j}$ and consequently subtracts
$S_{i}$ from current $S_{i}$, which results in a new status $S'_{i}=S_{i}\ominus S_{i}=0$.

Now, consider the change as it is experienced by $p_{j}$. Its agreement
with $p_{i}$ changes from $A_{i,j}$ to $A_{i,j}\oplus S_{i}$. However,
following from the triangle inequality and the convexity of $R$ (see
Section \ref{sub:Mathematical-notation}), it follows from $\overrightarrow{A_{i,j}}$
and $\overrightarrow{S_{i}}$ being in $R$ that the new agreement
$\overrightarrow{A'_{i,j}}=\overrightarrow{S_{i}\oplus A_{i,j}}$
is in $R$ as well. 

The status of $p_{j}$, $S_{j}$, also changes as a result of the
change in $X_{i,j}$. Since $X'_{i,j}=X_{i,j}\oplus S_{i}$, the new
value of $S_{j}$, $S'_{j}$, is $S'_{j}=S_{j}\oplus S_{i}$. Again,
since $\overrightarrow{S_{i}}$ and $\overrightarrow{S_{j}}$ are
in $R$, so is $\overrightarrow{S'_{j}}=\overrightarrow{S_{i}\oplus S_{j}}$.
Finally, for any other neighbor of $p_{j}$, $p_{k}\neq p_{i}\in N_{j}$,
the value of $S_{j}\ominus A_{j,k}$ is increased by $S_{i}$. Since
$\overrightarrow{S_{j}\ominus A_{j,k}}\in R$ and $\overrightarrow{S_{i}}\in R$,
$\overrightarrow{S'_{j}\ominus A_{j,k}}=\overrightarrow{S_{i}\oplus S_{j}\ominus A_{j,k}}\in R$. 

We conclude that when a leaf $p_{i}$ sends a message to its parent
$p_{j}$, $S_{i}$ becomes a zero element of $\mathcal{W}$ and the
conditions of the theorem continue to hold for $p_{i}$ and $p_{j}$.
The same happens when all of the leaves below any peer send their
message. Therefore, by induction, the same happens when a non-leaf
peer $p_{i}$ receives the final message from any of its descendants
and sends the message to its own parent $p_{j}$.

Finally, when the root of the spanning tree, call it $p_{0}$, has
received all of the messages, we have $S_{i}=0$ for any $p_{i}\neq p_{0}$
and $\overrightarrow{S_{0}}\in R$. From Theorem \ref{lem:The-global-sum},
$\bigoplus X=\bigoplus_{p_{i}\in P}S_{i},$ and since in $\bigoplus_{p_{i}\in P}S_{i}=S_{0}\oplus S_{1}\oplus\dots\oplus S_{n}$
the latter $n-1$ elements are zero elements, we have that $\bigoplus X=\bigoplus_{p_{i}\in P}S_{i}=S_{0}$.
Thus, $\overrightarrow{\bigoplus X}=\overrightarrow{S_{0}}\in R$,
which proves the theorem.
\end{IEEEproof}
Relying on Theorem \ref{thm:main}, peers can indeed stop messages
once they compute that the conditions of the local stopping rule (Definition
\ref{[Local-stopping-rule]}) are satisfied. If some $p_{i}$ stops
sending messages according to the local stopping rule then there are
two possible cases: The first is that $\overrightarrow{S_{i}}$ is
in the same region as $\overrightarrow{\bigoplus X}$ and hence $p_{i}$
computes the correct outcome. The other is that there are some peers
in the network for whom the conditions of the local stopping rule
are not satisfied. In the latter case, those peers are guaranteed
to continue sending messages until the outcome of $p_{i}$ is corrected.

\section{\label{sec:Balance-correction}Balance correction}

A local stopping rule is just one part of a local thresholding algorithm.
It must be complemented by a method for achieving the conditions set
by the local stopping rule. In this section we first prove that for
any set of inputs, any network topology, and any convex region $R$
containing $\bigoplus X$, there exists a solution which satisfies
the conditions of the local stopping rule (Def. \ref{[Local-stopping-rule]}).
Then, we describe a pair of local correction policies, each of which
is a formula by which a peer can compute outgoing messages such that
after the messages are sent the conditions hold at that peer.

\subsection{Existence of a solution}

We first show that regardless of the topology and the input of the
different peers, there is a set of values for the different $X_{i,j}$
such that all non-zero $\overrightarrow{S_{i}\ominus A_{i,j}}$ are
equal to all non-zero $\overrightarrow{A_{i,j}}$. This means that
they all must reside in the same $R$, regardless of its shape. Consider
any spanning tree of the network graph. The solution is to assign
zero weight to $X_{i,j}$ and $X_{j,i}$ corresponding to edges not
on a spanning tree. The other $X_{i,j}$ and $X_{j,i}$ will be assigned
values which provide that for every $p_{i}\in P$ and $p_{j}\in N_{i}$
the following holds: $S_{i}\ominus A_{i,j}=A_{i,j}=\frac{1}{2\left|V\right|}\odot\bigoplus X$.
\begin{thm}
\label{lem:existence}{[}Termination state existence{]} For any connected
network graph and any set of inputs and a convex $R$ such that $\overrightarrow{\bigoplus X}\in R,$
there is a setup of values for $X_{i,j}$ such that the conditions
of the local stopping rule are met.\end{thm}
\begin{IEEEproof}
Let $T\left(V,E'\right)$ be a spanning tree over $G\left(V,E\right)$
and let $V_{i}$ be the vertices in $p_{i}$'s subtree. Let the global
weighted average be $\bigoplus X$. For every $\left(p_{i},p_{k}\right)\in E$
such that $\left(p_{i},p_{k}\right)\notin E',$ let $\left|X_{i,k}\right|=\left|X_{k,i}\right|=0,$
which satisfies the conditions of the local stopping rule.

Define the subtree status $Y_{i}$ recursively as follows: For a leaf
node, the subtree status is equal to the input, $Y_{i}=X_{i}$. For
a non-leaf node, $Y_{i}$ is the status omitting the last message
sent to and received from its parent, $Y_{i}=X_{i,i}\oplus\left(\bigoplus_{p_{k}\neq p_{j}\in N_{i}}X_{k,i}\ominus X_{i,k}\right)$. 

Now, we define a message which should be sent and one which should
be received from each node to any neighbor on the spanning tree. Let
$p_{i}$ be a node and $p_{j}$ its parent on the tree. We set $X_{i,j}=\frac{1}{2}\odot Y_{i}\ominus\frac{1}{4\left|V\right|}\odot\bigoplus X$
and $X_{j,i}=\frac{3}{4\left|V\right|}\odot\bigoplus X\ominus\frac{1}{2}\odot Y_{i}$.

If $p_{i}$ is a leaf, then $X_{i,j}=\frac{1}{2}\odot Y_{i}\ominus\frac{1}{4\left|V\right|}\odot\bigoplus X=\frac{1}{2}\odot X_{i}\ominus\frac{1}{4\left|V\right|}\odot\bigoplus X$
and $X_{j,i}=\frac{3}{4\left|V\right|}\odot\bigoplus X\ominus\frac{1}{2}\odot X_{i}$.
It follows that $S_{i}\ominus A_{i,j}=A_{i,j}=X_{i,j}\oplus X_{j,i}=\frac{1}{2\left|V\right|}\odot\bigoplus X$,
and that $S_{i}=2\odot A_{i,j}=\frac{1}{\left|V\right|}\odot\bigoplus X$.
Thus, the vector parts of $S_{i}$, of $A_{i,j}$ and of $S_{i}\ominus A_{i,j}$
are all equal to $\overrightarrow{\bigoplus X}$. Additionally, $X_{i,j}\ominus X_{j,i}=X_{i}\ominus\frac{1}{\left|V\right|}\odot\bigoplus X$. 

Next, consider a peer $p_{i}$ connected to $\left|N_{i}\right|-1$
leaves and to a parent $p_{j}$. Thus, $p_{i}$ has $Y_{i}=X_{i,i}\oplus\left(\bigoplus_{p_{k}\neq p_{j}\in N_{i}}\left(X_{k,i}\ominus X_{i,k}\right)\right)=X_{i,i}\oplus\left(\bigoplus_{p_{k}\neq p_{j}\in N_{i}}\left(X_{k}\ominus\frac{1}{\left|V\right|}\odot\bigoplus X\right)\right)$.
Hence, $p_{i}$ computes with its own parent $p_{j}$ $X_{i,j}\ominus X_{j,i}=Y_{i}\ominus\frac{1}{\left|V\right|}\odot\bigoplus X=\bigoplus_{p_{k}\in V_{i}}X_{k,k}\ominus\frac{1}{\left|V\right|}\odot\bigoplus X$,
which is the difference between the average of the inputs in $p_{i}$'s
subtree and the global average. Applying induction, we get that for
any $p_{i}$ and its parent $p_{j}$ we have $Y_{i}=X_{i,i}\oplus\left(\bigoplus_{p_{k}\neq p_{i}\in N_{i}}\left(X_{k,i}\ominus\frac{\left|V_{k}\right|}{\left|V\right|}\odot\bigoplus X\right)\right)=X_{i,i}\oplus\left(\bigoplus_{p_{k}\neq p_{i}\in V_{i}}\left(X_{k,k}\ominus\frac{1}{\left|V\right|}\odot\bigoplus X\right)\right)$
and $X_{i,j}\ominus X_{j,i}=\left(\bigoplus_{p_{k}\in V_{i}}X_{k,k}\right)\ominus\frac{\left|V_{i}\right|}{\left|V\right|}\odot\bigoplus X$. 

It follows that for every peer $p_{i}$ and any $p_{j}$ (parent or
non-parent), we have $A_{i,j}=S_{i}\ominus A_{i,j}=\frac{1}{2\left|V\right|}\odot\bigoplus X$,
and thus $\overrightarrow{A_{i,j}}=\overrightarrow{S_{i}\ominus A_{i,j}}=\frac{1}{2\left|V\right|}\odot\overrightarrow{\bigoplus X}$,
which agrees with the conditions of the local stopping rule regardless
of what $R$ is.
\end{IEEEproof}
We note, once more, that although the proof of Theorem \ref{lem:existence}
uses a spanning tree, the correctness of the stopping rule does not
rely on the topology. The proof only serves to show that at least
one global termination state exists.

However, a peer cannot directly compute this specific termination
state unless it has knowledge of the global topology and of all inputs.
Therefore, we now move to describe how local correction methods can
be directly computed by a peer using only its own state and agreements
with neighbors. These methods do not necessarily provide global termination.
Rather, they restore the stopping condition at the peer which uses
them, while possibly by violating the condition of its neighbor.

\subsection{\label{sub:Local-heuristics}Local correction}

Consider a peer $p_{i}$ whose state currently violates the conditions
of the local stopping rule, i.e., for some of the $p_{j}\in N_{i,}$
either $\overrightarrow{A_{i,j}}\not\in R$ and $\left|A_{i,j}\right|>0$
or $\overrightarrow{S_{i}\ominus A_{i,j}}\not\in R$ and $\left|S_{i}\ominus A_{i,j}\right|>0$.
A local correction heuristic is a method for computing messages for
some of the peers in $N_{i}$ such that the new values of the respective
$X_{i,j}$'s cause the conditions of the local stopping rule to hold
for $p_{i}$. Since this paper describes a general algorithm, we are
interested in local correction methods whose success does not depend
on the state of the peer, the topology, and the specific $R$ in question. 

A simple way to achieve independence of $R$ is to make sure that
after the messages are sent, all $\overrightarrow{A_{i,j}}$ and all
$\overrightarrow{S_{i}\ominus A_{i,j}}$ are equal. This way, if one
is inside $R$, then all are inside $R$ and the conditions of the
local stopping rule are met. First, note that $\overrightarrow{S_{i}\ominus A_{i,j}}=\overrightarrow{A_{i,j}}$
means $\overrightarrow{S_{i}}=\overrightarrow{A_{i,j}\oplus A_{i,j}}=\overrightarrow{2\odot A_{i,j}}=\overrightarrow{A_{i,j}}$,
i.e., the local correction method must compute new values for the
$X_{i,j}$ of $p_{j}\in N_{i}$ such that 
\begin{equation}
\forall p_{j}\in N_{i}:\overrightarrow{A_{i,j}}=\overrightarrow{S_{i}}.\label{eq:requirement}
\end{equation}

\begin{thm}
\label{thm:General-A'ij}{[}Perfect correction{]} Let $A'_{i,j}$
and $S'_{i}$ denote the new values computed by changing each $X_{i,j}$
to $X'_{i,j}$. Then $\overrightarrow{A'_{i,j}}=\overrightarrow{S'_{i}}$
for all $p_{j}\in N_{i}$ if and only if $A'_{i,j}=\frac{\left|A'_{i,j}\right|}{\left|X_{i,i}\oplus\bigoplus_{p_{k}\in N_{i}}2\odot X_{k,i}\right|}\odot\left(X_{i,i}\oplus\bigoplus_{p_{k}\in N_{i}}2\odot X_{k,i}\right)$.\end{thm}
\begin{IEEEproof}
Because the different $\overrightarrow{A'_{i,j}}$ are all equal to
$\overrightarrow{A'_{i,i}}$, we write for all $p_{j},p_{k}\in N_{i}$:
\begin{eqnarray*}
\overrightarrow{A'_{i,j}} & = & \overrightarrow{A'_{i,k}}\\
\overrightarrow{X'_{i,j}\oplus X_{j,i}} & = & \overrightarrow{X'_{i,k}\oplus X_{k,i}}\\
\overrightarrow{X'_{i,k}} & = & \overrightarrow{\left(X'_{i,j}\oplus X_{j,i}\right)\ominus X_{k,i}}\\
\overrightarrow{X_{k,i}\ominus X'_{i,k}} & = & \overrightarrow{\left(2\odot X_{k,i}\right)\ominus\left(X'_{i,j}\oplus X_{j,i}\right)}.
\end{eqnarray*}

The equation $\forall p_{j}\in N_{i}:\overrightarrow{A'_{i,j}}=\overrightarrow{S'_{i}}$
can be rewritten
\[
\forall p_{j}\in N_{i}:\overrightarrow{X'_{i,j}\oplus X_{j,i}}=\overrightarrow{X_{i,i}\oplus\bigoplus_{p_{k}\in N_{i}}\left(X_{k,i}\ominus X'_{i,k}\right)}
\]
\[
=\overrightarrow{X_{i,i}\oplus\bigoplus_{p_{k}\in N_{i}}\left[\left(2\odot X_{k,i}\right)\ominus\left(X'_{i,j}\oplus X_{j,i}\right)\right]}
\]
\[
=\overrightarrow{X_{i,i}\oplus\bigoplus_{p_{k}\in N_{i}}\left[2\odot X_{k,i}\right]\ominus\bigoplus_{p_{k}\in N_{i}}\left(X'_{i,j}\oplus X_{j,i}\right)}
\]
\[
=\overrightarrow{X_{i,i}\oplus\bigoplus_{p_{k}\in N_{i}}\left[2\odot X_{k,i}\right]\ominus\left[\left|N_{i}\right|\odot\left(X'_{i,j}\oplus X_{j,i}\right)\right]},
\]

which is equivalent to stating, $\forall p_{j}\in N_{i}$, that:
\begin{eqnarray*}
\overrightarrow{\left(\left|N_{i}\right|+1\right)\odot\left(X'_{i,j}\oplus X_{j,i}\right)} & = & \overrightarrow{X_{i,i}\oplus\bigoplus_{p_{k}\in N_{i}}\left[2\odot X_{k,i}\right]}.
\end{eqnarray*}

Since multiplication by a constant only changes the weight and not
the vector part of a weighted vector,we have that 
\begin{equation}
\forall p_{j}\in N_{i}:\overrightarrow{X'_{i,j}\oplus X_{j,i}}=\overrightarrow{X_{i,i}\oplus\bigoplus_{p_{k}\in N_{i}}\left[2\odot X_{k,i}\right]}.\label{eq:General-Aij-1-1}
\end{equation}

The set of possible $A'_{i,j}$ which satisfy the requirement in Eq.
\ref{eq:requirement} can be computed by normalizing the weighted
vector to the desired $\left|A'_{i,j}\right|$: 
\begin{equation}
A'_{i,j}=\frac{\left|A'_{i,j}\right|}{\left|X_{i,i}\oplus\bigoplus_{p_{k}\in N_{i}}2\odot X_{k,i}\right|}\odot\left(X_{i,i}\oplus\bigoplus_{p_{k}\in N_{i}}2\odot X_{k,i}\right).\label{eq:General-A'ij}
\end{equation}

\end{IEEEproof}
In other words, what Theorem \ref{thm:General-A'ij} states is that
if the peer chooses a new weight for its agreement with a neighbor,$\left|A'_{i,j}\right|$,
then that weight dictates the value of $\overrightarrow{A'_{i,j}}$
which would satisfy Eq. \ref{eq:requirement}. Since the peer can
enforce both $\left|A'_{i,j}\right|$ and $\overrightarrow{A'_{i,j}}$
by choosing appropriate $\left|X'_{i,j}\right|$ and $\overrightarrow{X'_{i,j}}$,
Theorem \ref{thm:General-A'ij} identifies a range of outgoing messages
that satisfy this Eq. \ref{eq:requirement}, which are those that
can be computed using Eq. \ref{eq:General-A'ij}.

\subsection{Weight distribution schemes}

Theorem \ref{thm:General-A'ij} provides the general solution to the
problem of computing outgoing messages that will set $p_{i}$'s state
to one which obeys the conditions of the local stopping rule. However,
the equation does not dictate a single solution but rather a proportion
between the weight of the outgoing message, $\left|X'_{i,j}\right|$,
and the vector of that message, $\overrightarrow{X'_{i,j}}$. Each
choice of $\left|X'_{i,j}\right|$ will have a different effect on
the states of the sender and the recipient.

The most important property to be had in a correction scheme is guaranteed
convergence to a global solution. From results obtained for other
iterative averaging algorithms \cite{Gossip}, we know that it is
simple to achieve convergence of the $\left|S_{i}\right|$ of all
$p_{i}$ to a constant value that depends only on the topology: each
time a peer's state violates the stopping criterion, it will distribute
half of $\left|S_{i}\right|$ to its neighbors. So the total weight
to be distributed is given by the equation 
\begin{equation}
\sum_{p_{j}\in N_{i}}\left(\left|X'_{i,j}\right|-\left|X_{i,j}\right|\right)=\frac{\left|S_{i}\right|}{2}.\label{eq:HalfAii}
\end{equation}
However, naive implementation of this policy may lead to very small
$\left|S_{i}\right|$ values at some peers. While mathematically small
weights do not pose a problem, in practice they would lead to numeric
instability. Therefore, a minimal weight $\beta$ is enforced on $\left|S_{i}\right|$
by using $\sum_{p_{j}\in N_{i}}\left(\left|X'_{i,j}\right|-\left|X_{i,j}\right|\right)=\frac{\left|S_{i}\right|-\beta}{2}$.
We find that a small $\beta$ does not hinder convergence.

\subsubsection{Uniform weight distribution}

A simple weight distribution policy is to allocate a constant portion
of $\left|S_{i}\right|$ to each $p_{j}\in N_{i}$, i.e., to set $\left|X'_{i,j}\right|-\left|X_{i,j}\right|=\frac{\left|S_{i}\right|-\beta}{2\left|N_{i}\right|}$
. Since $\left|X'_{i,j}\right|-\left|X_{i,j}\right|=\left|A'_{i,j}\right|-\left|A_{i,j}\right|$,
the new $\left|A'_{i,j}\right|=\frac{\left|S_{i}\right|-\beta}{2\left|N_{i}\right|}+\left|A_{i,j}\right|$.
We denote this the \emph{uniform weight distribution }method and formally
define it by instantiating Eq. \ref{eq:General-A'ij} to:
\begin{equation}
A'_{i,j}=\frac{\left|A_{i,j}\right|+\frac{\left|S_{i}\right|-\beta}{2\left|N_{i}\right|}}{\left|X_{i,i}\oplus\bigoplus_{p_{k}\in N_{i}}2\odot X_{k,i}\right|}\odot\left(X_{i,i}\oplus\bigoplus_{p_{k}\in N_{i}}2\odot X_{k,i}\right).\label{eq:UniformWeightDistribution}
\end{equation}

\subsubsection{Selective local correction}

By distributing the weight uniformly, as described in Eq. \ref{eq:UniformWeightDistribution}
above, a new value is computed for every $X_{i,j}$. This is often
unnecessary as many of the neighbors $p_{j}\in N_{i}$ may have $\overrightarrow{A_{i,j}}$
and $\overrightarrow{S_{i}\ominus A_{i,j}}$ that still fall inside
$R$. Setting the $X_{i,j}$ of those neighbors to a new value can
be doubly wasteful: a message must be sent to every neighbor $p_{j}$,
and the change of $X_{i,j}$ might well change some $S_{j}\ominus A_{j,k}$
at $p_{j}$ to the degree that it violates the stopping condition,
triggering further messages. A solution which selectively sends messages
to only part of the neighbor set $N_{i}$ and still brings all of
the $A_{i,j}$ and $S_{i}\ominus A_{i,j}$ into $R$ is therefore
desirable.

Denote $V_{i}$ the set of neighbors for whom the stopping condition
is violated, $V_{i}=\left\{ p_{j}\in N_{i}:\overrightarrow{A_{i,j}}\notin R\vee\overrightarrow{S_{i}\ominus A_{i,j}}\notin R\right\} $.
The complementary set, $N_{i}\setminus V_{i}$, are the neighbors
whose update may be avoided. Consider an imaginary peer $p_{im}$
with $N_{im}=V_{i}$, $X_{im,j}=X_{i,j}$ and $X_{j,im}=X_{j,i}$
for all $p_{j}\in N_{i}$, and $X_{im,im}=X_{i,i}\oplus\bigoplus_{p_{j}\in N_{i}\setminus V_{i}}X_{j,i}\ominus X_{i,j}$.
Note that $X_{im,im}\oplus\bigoplus_{p_{j}\in N_{im}}2\odot X_{j,im}=S_{im}\oplus\bigoplus_{p_{j}\in N_{im}}A_{im,j}=S_{i}\oplus\bigoplus_{p_{j}\in V_{i}}A_{i,j}$.
If $p_{im}$ sets every $X_{im,j}$ according to Eq. \ref{eq:General-Aij-1-1},
then we have
\begin{eqnarray}
A'_{im,j}\\
= & \frac{\left|A'_{im,j}\right|}{\left|S_{im}\oplus\bigoplus_{p_{k}\in N_{im}}A_{im,k}\right|}\odot\left(S_{im}\oplus\bigoplus_{p_{k}\in N_{im}}A_{im,k}\right)\\
= & \frac{\left|A'_{i,j}\right|}{\left|S_{i}\oplus\bigoplus_{p_{k}\in V_{i}}A_{i,k}\right|}\odot\left(S_{i}\oplus\bigoplus_{p_{k}\in V_{i}}A_{i,k}\right)\label{eq:shorter}\\
= & \frac{\left|A'_{i,j}\right|\odot\left(X_{i,i}\oplus\bigoplus_{p_{k}\in N_{i}\setminus V_{i}}\left(X_{k,i}\ominus X_{i,k}\right)\oplus\bigoplus_{p_{k}\in V_{i}}2\odot X_{k,i}\right)}{\left|X_{i,i}\oplus\bigoplus_{p_{k}\in N_{i}\setminus V_{i}}\left(X_{k,i}\ominus X_{i,k}\right)\oplus\bigoplus_{p_{k}\in V_{i}}2\odot X_{k,i}\right|} & .\label{eq:longer}
\end{eqnarray}
It then would follow that if for all $p_{j}\in N_{im}$ (i.e., in
$V_{i}$), $A'_{im,j}$ (which is equal to $A'_{i,j}$) is set according
to the policy detailed in Eq. \ref{eq:UniformWeightDistribution},
then $\left|S'_{im}\right|=\left|S'_{i}\right|=\frac{\left|S_{i}\right|-\beta}{2}$.
The selective version of the uniform weight distribution policy is
therefore:
\begin{equation}
A'_{i,j}=\frac{\left|A_{i,j}\right|+\frac{\left|S_{i}\right|-\beta}{2\left|V_{i}\right|}}{\left|S_{i}\oplus\bigoplus_{p_{k}\in V_{i}}A_{i,k}\right|}\odot\left(S_{i}\oplus\bigoplus_{p_{k}\in V_{i}}A_{i,k}\right).\label{eq:selectiveuniformcorrection}
\end{equation}

The problem with a selective policy lies with the neighbors that did
not violate the stopping condition, those in $N_{i}\setminus V_{i}$.
Since setting the $X_{i,k}$ of each $p_{k}\in V_{i}$ changes the
status $S_{i}$, it may well be that for $p_{j}\in N_{i}\setminus V_{i}$
the vector $\overrightarrow{S_{i}\ominus A_{i,j}}$ is no longer in
$R$. One solution can be to iteratively add neighbors to $V_{i}$
if they violate the stopping condition, and to terminate the iteration
when $V_{i}$ no longer grows. At this stage only, messages are sent
to all of the neighbors in $V_{i}$. Note that, at worst, iteration
ends with $V_{i}=N_{i}$, which is the non-selective solution.

\section{\label{sec:Source-selection}Source selection}

The stopping and correction methods presented in the previous sections
are general. They can be applied to various sets of convex regions.
In this section we demonstrate their application to the problem of
source selection. The source selection problem is a generalization
of the majority voting problem in which votes and options are vectors
in $\mathbb{R}^{d}$ rather than the points $\left\{ 0,1\right\} $.
The generalization is sufficiently rich to allow reduction from data
mining problems such as decision tree induction \cite{p2pDT} and
$k$-median \cite{FLP}, and yet simple enough to allow thorough and
application-independent experimentation.

Let $C=\left\{ c_{1},c_{2},\dots,c_{k}\in\mathbb{R}^{d}\right\} $
be a set of options and let $X_{i,i}$ be the input of $p_{i}$ such
that $\left|X_{i,i}\right|=1$ and $\overrightarrow{X_{i,i}}\in\mathbb{R}^{d}$.
The objective of the peers in the source selection problem is to compute
$f\left(\overrightarrow{\bigoplus X}\right)=\arg\min_{c_{i}\in C}\left\{ \left\Vert c_{i}-\overrightarrow{\bigoplus X}\right\Vert \right\} $,
where the norm $\left\Vert \cdot\right\Vert $ can be any norm (we
use the L2 norm). Note that $\mathcal{R}=\left\{ \left[c_{1}\right],\left[c_{2}\right],\dots,\left[c_{k}\right]\right\} $,
where $\left[c_{i}\right]=\left\{ x\in\mathbb{R^{d}}:f\left(x\right)=c_{i}\right\} $
is a set of convex regions, and $nil$, the complementary region,
is empty%
\footnote{To see that this problem reduces to a majority vote, consider $\overrightarrow{X_{i,i}}\in\left\{ 0,1\right\} $
and $C=\left\{ 0,1\right\} $.%
}. Thus, we can apply local thresholding to the problem with the new
stopping rule and the local correction policies described in Section
\ref{sub:Local-heuristics}. 

To solve the source selection problem, peers follow the general setup
presented above. They exchange weighted vectors and retain the latest
vector sent and the latest received from each neighbor. They maintain
and update $A_{i,j}$ and $S_{i}$ with every incoming message and
every change in their input, and they react to violations of the stopping
rule by computing corrective messages.

To this general framework, the local source selection algorithm for
general network graphs (LSS, for short) (Alg. \ref{alg:Mean-Selection})
makes three modifications. First, it always evaluates the stopping
rule with respect to the convex region $f\left(\overrightarrow{S_{i}}\right)$.
Second, it attaches a sequential number to every outgoing message,
so that the recipient can ignore late arrivals. This is because, in
a real system and in our simulations, messages do not necessarily
arrive in the order in which they were sent. Finally, it places a
strict lower bound of $\ell$ time units between subsequent outgoing
messages, which is necessary in order to control the number of events. 

\begin{algorithm}
\caption{\label{alg:Mean-Selection}Local Source Selection in General Network
Graphs}

\textbf{Common inputs for all peers:} $\beta\in\left[0,1\right]$,
$\ell\in\mathbb{N}$, $C=\left\{ c_{0},c_{1},\dots,c_{k}\in\mathbb{R}^{d}\right\} $

\textbf{Private input of $p_{i}$:} $\overrightarrow{x_{i}}\in\mathbb{R}^{d}$,
$N_{i}\subset V$

\textbf{Output of $p_{i}$: $f\left(\overrightarrow{S_{i}}\right)$}

\textbf{Initialization: }

$X_{i,i}\leftarrow\left(\overrightarrow{x_{i}},1\right)$, $\ell_{i}\leftarrow-\ell$,
$seq_{i}\leftarrow0$

For all $p_{j}\in N_{i}$ 

- $X_{i,j},X_{j,i}\leftarrow\left(\overline{0},0\right)$

- $last_{j}\leftarrow0$

\textbf{On a message $\left\langle X,seq\right\rangle $ from $p_{j}\in N_{i}$:}

If $seq\geq last_{j}$ then 

- $last_{j}\leftarrow seq$

- $X_{j,i}\leftarrow X$

- $A_{i,j}\leftarrow X_{i,j}\oplus X_{j,i}$

- $S_{i}\leftarrow X_{i,i}\oplus\bigoplus_{p_{j}\in N_{i}}\left(X_{j,i}\ominus X_{i,j}\right)$

\textbf{On initialization, on any change to $S_{i}$, and on timer
expiration:} 

If $currentTime()-\ell_{i}<\ell$

- Set timer to $currentTime()+\ell-\ell_{i}$ and return

{\footnotesize $V_{i}\leftarrow\left\{ p_{j}\in N_{i}:f\left(\overrightarrow{A_{i,j}}\right)\neq f\left(\overrightarrow{S_{i}}\right)\vee f\left(\overrightarrow{A_{i,i}\ominus A_{i,j}}\right)\neq f\left(\overrightarrow{S_{i}}\right)\right\} $}{\footnotesize \par}

If $V_{i}=\emptyset$ return

$oldS_{i}\leftarrow S_{i}$

Do 

- $newS{}_{i}\leftarrow oldS_{i}\oplus\bigoplus_{p_{j}\in V_{i}}A_{i,j}$

- $\forall p_{j}\in V_{i}$ do $X_{i,j}\leftarrow\left(\frac{\frac{\left|oldS_{i}\right|-\beta}{2\left|V_{i}\right|}+\left|A_{i,j}\right|}{\left|newS_{i}\right|}\odot newS_{i}\right)\ominus X_{j,i}$

- $S_{i}\leftarrow X_{i,i}\oplus\bigoplus_{p_{j}\in N_{i}}\left(X_{j,i}\ominus X_{i,j}\right)$

- {\footnotesize $W_{i}\leftarrow\left\{ p_{j}\in N_{i}:f\left(\overrightarrow{A_{i,j}}\right)\neq f\left(\overrightarrow{S_{i}}\right)\vee f\left(\overrightarrow{A_{i,i}\ominus A_{i,j}}\right)\neq f\left(\overrightarrow{S_{i}}\right)\right\} $}{\footnotesize \par}

- $V_{i}\leftarrow V_{i}\cup W_{i}$

While $W_{i}\neq\emptyset$

$seq_{i}\leftarrow seq_{i}+1$

$\ell_{i}\leftarrow currentTime()$

For all $p_{j}\in V_{i}$ send $\left\langle X_{i,j},seq_{i}\right\rangle $
to $p_{j}$
\end{algorithm}

\section{\label{sec:Experimental-evaluation}Experimental validation}

The main contribution of this work is not a new algorithm for a specific
data mining problem but rather a fundamental change which can be applied
to all existing local thresholding algorithms. The purpose of the
following section is to evaluate the effect of the main arguments
of the problem -- the system and its dynamics, the input domain $\mathcal{V}$
and the output domain $\mathcal{R}$, and the algorithm parameters
-- on performance.

To make the evaluation more specific, it is carried out in the context
of the LSS algorithm. However, performance on specific applications
and in specific system settings would have to be evaluated per case.
To facilitate that, the experiments were carried out with a standard
simulator, peersim \cite{peersim}, and the code is available on-line.

\subsection{Experimental setup}

LSS performance is influenced by four types of parameters. The first
are those of the system: the number of peers, $n$, the topology in
which they are arranged, and the message drop rate, $r$. In our experiments
we use three different topologies. To investigate unstructured peer-to-peer
systems we use the Barabasi-Albert \cite{BA} model, which is a well-known
approximation of the Internet router topology and which is claimed
to approximate the structure of systems like Gnutella \cite{gnutella}.
To investigate structured peer-to-peer systems, we use the popular
Chord topology \cite{chord} with the variant that connection with
fingers is assumed to be bidirectional (in essence, implementing Symmetric
Chord \cite{SChord}). For the third target topology, wireless sensor
networks, no standard accepted model exists, and mere connectedness
is a design challenge \cite{Li2009}. We therefore opt for a wireless
sensor network in which sensors are locations on a bi-dimensional
grid. In all those three topologies, the average connectivity $\left|N_{i}\right|$
can be controlled, although in Chord its value is typically much larger
($\log\left(n\right)$) than in the other two. We experiment with
reliable communication and with a range of message drop rates.

The second type of parameters are those related to the function computed,
which in the case of LSS are the number of different sources from
which the selection is made, $k$, the dimensionality of the data,
$d$. and the distribution of the data. In our experiments, the data
is normally and independently distributed along each dimension. We
randomly select one of the possible sources as the desired outcome
of the algorithm and denote the contender its nearest neighbor. The
mean of the data is set to a weighted average of the desired outcome
and the contender. The weight given to the contender, between zero
and one-half, is denoted the \emph{bias} of the data. The standard
deviation of the data is selected as a multiplier of the distance
between the desired outcome and the contender. That multiplier is
denoted \emph{std}. When the data is dynamic, inputs are resampled
from the same distribution at every cycle. The proportion of peers
whose data changes at every simulation cycle is denoted the\emph{
noise rate,} which is measured in units of changed peers per million
simulation cycles (ppmc).

\begin{figure}
\caption{\label{fig:Example-of-data}Example of data}

\includegraphics[width=0.9\columnwidth]{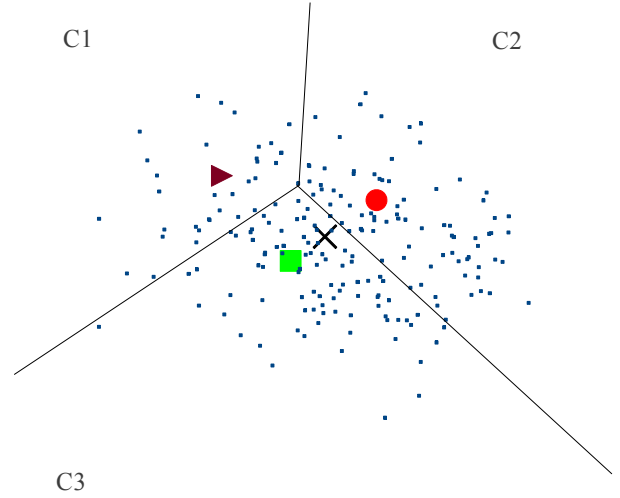}
\end{figure}

Figure \ref{fig:Example-of-data} depicts an example of two hundred
data points with $d=2$, $k=3$. The mean of the data is denoted by
the X, which is located at a bias of 40\% between the desired outcome
(square) and the contender (circle). The standard deviation in this
example equals once the distance between the desired outcome and the
contender ($std=100\%$).

The last type of parameters are those which can be controlled by the
user: the minimum weight parameter $\beta$ and the lower bound on
the delay between subsequent messages, $\ell$.

Four sets of experiments were carried out. The first two survey the
effect of each parameter using static data, the third examines dynamic
data and the last the effect of system dynamics. In each experiment,
all but one of the parameters were set to a default value: $n=10,000$,
$\left|N_{i}\right|\simeq4$, a drop rate of zero (i.e., reliable
communication), $k=3$, $d=2$, a bias of 10\% and std of 100\%, $\beta=0.001$
and $\ell=1$. Then, the simulation was run ten times for each tested
value of the remaining parameter. 

In all of the experiments in which the data is static, performance
is measured by the number of cycles needed for convergence of 95\%
and 100\% of the peers, and by the number of messages required for
convergence of all peers (convergence of the last 5\% of the peers
usually requires very few messages). To allow comparison of different
topologies, the average number of messages per edge is reported, rather
than the grand total. When the data is dynamic, convergence never
occurs and messages never cease. The performance metrics used are,
therefore, the average accuracy (percent of peers computing the wrong
outcome) and average communication cost (number of messages sent per
communication link per cycle). In the literature this last metric
is sometimes called normalized messaging and has a maximal value of
two in case $\ell=1$ and $\frac{2}{\ell}$ in general.

\subsection{\label{sub:System-properties}System properties}

\begin{figure*}
\subfloat[\label{fig:scaleup-util-uniform}Cycles to convergence]{\includegraphics[width=0.5\textwidth]{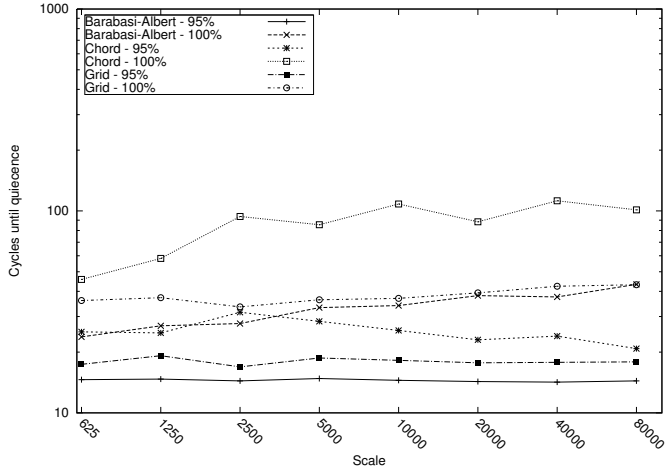}}\subfloat[\label{fig:scaleup-msgs-uniform}Messages per cycle per connection]{\includegraphics[width=0.5\textwidth]{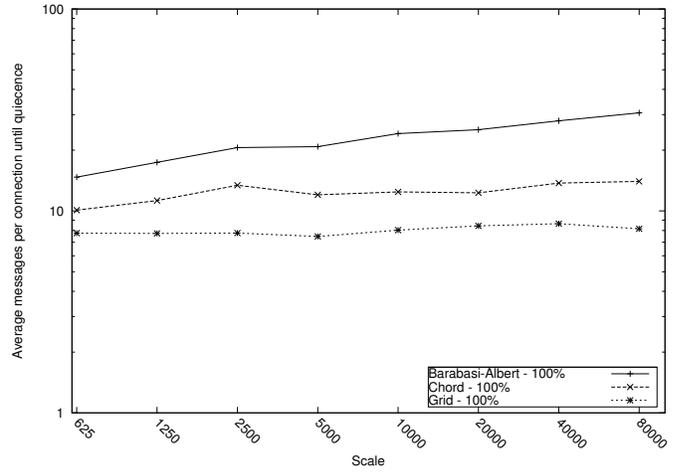}}\caption{\label{fig:Scaleup-of-uniform}Scale-up}
\end{figure*}

For every distributed algorithm, and especially one intended for large
systems, scalability is the single most important criterion. Figure
\ref{fig:Scaleup-of-uniform} depicts the scale-up of LSS with selective
uniform correction. The number of cycles required for complete quiescence
and for convergence of 95\% of the peers is depicted on the left (sub-figure
\ref{fig:scaleup-util-uniform}), and the number of messages per link
is depicted on the right (sub-figure \ref{fig:scaleup-msgs-uniform}.)

The first thing to notice is that LSS overhead seems to converge to
a constant as the system is scaled up. This is certainly true for
the number of cycles until 95\% of the peers converge, and for the
communication. Although convergence of 100\% percent of the peers
is an interesting metric, its value is mainly theoretical. First,
it is a worst-case metric that depends on the worst performing peer.
Second, the typical working scenario of a large distributed system
is dynamic, and does not allow 100\% convergence at all.

\begin{figure*}
\subfloat[\label{fig:Ni-util} Effect of average$\left|N_{i}\right|$ on convergence]{\includegraphics[width=0.5\textwidth]{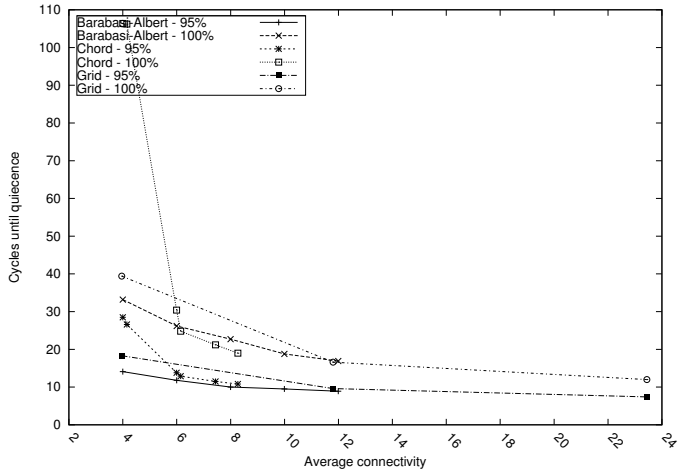}}\subfloat[\label{fig:Ni-msgs}Effect of average$\left|N_{i}\right|$ on communication]{\includegraphics[width=0.5\textwidth]{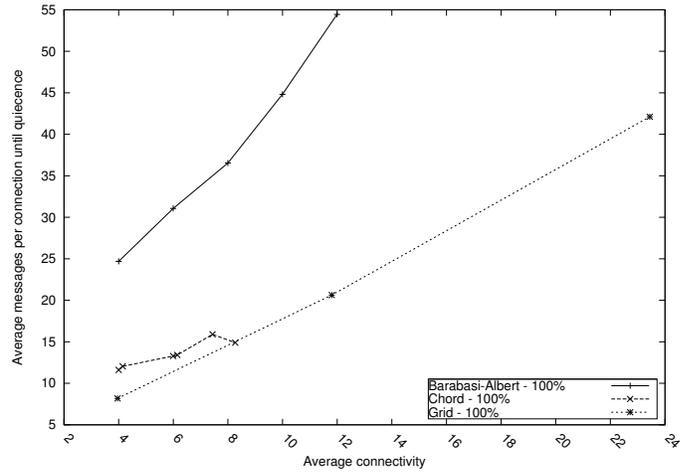}}

\caption{Connectivity}
\end{figure*}

The second thing to notice is the instability of the performance when
the topology is Barabasi-Albert. A deeper look into the results reveals
that performance is greatly influenced by outliers: single experiments
in which the overhead was exceptionally high. We note that Barabasi-Albert
is different from both Chord and Grid topologies in the sense that
there is no strict limit on peer connectivity. As can be seen in Figure
\ref{fig:Ni-msgs}, Barabasi-Albert is also the most sensitive to
average peer connectivity. Since each experiment is carried out using
a constant topology, topological effects are not averaged out and
may well explain outliers. 

\begin{figure*}
\subfloat[\label{fig:loss-util}Cycles to convergence]{\includegraphics[width=0.5\textwidth]{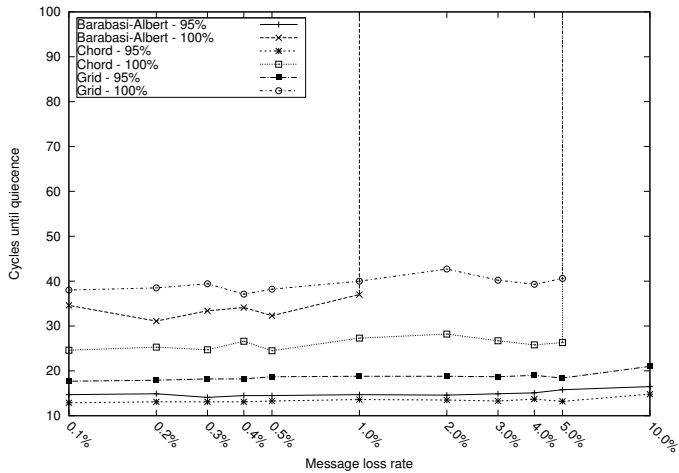}}\subfloat[\label{fig:loss-msgs}Messages per cycle per connection]{\includegraphics[width=0.5\textwidth]{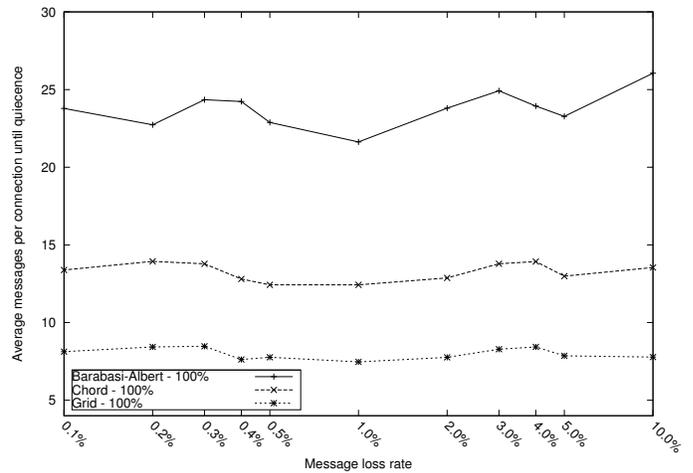}}\caption{Message loss rate}
\end{figure*}

Besides scale, the other important property of the system is connectivity,
as measured in the average size of $\left|N_{i}\right|$. Because
of the inert differences between the three topologies tested, not
all were tested on the same range of average $\left|N_{i}\right|$.
However, as can be seen in figures \ref{fig:loss-util} and \ref{fig:loss-msgs},
the effect of increased connectivity on LSS is to expedite convergence
and increase the number of messages per connection. Since the increase
in communication load per link appears to be linear while the number
of required converges quickly to a constant, there seems to be an
optimum point, which in this experiment is around $\left|N_{i}\right|=6$.
This could be an important observation because many systems do allow
at least limited control of connectivity.

The third and last important feature of the system is message reliability.
In Internet based systems, reliable messaging usually costs very little.
Even when reliability is not possible, message loss rate is expected
to be very low. Wireless sensor networks are drastically different:
message loss rates can be expected to be very high even between immediate
neighbors and reliable messaging is usually too costly to be used
for intensive computation. 

Figures \ref{fig:loss-util} and \ref{fig:loss-msgs} depict the effect
of random and independent message loss on the convergence and the
message overhead of LSS. As can be seen, limited message loss has
no impact on convergence or messaging overhead. This is attributed
to the effect of having multiple paths between peers: so long as corrective
messages arrive through one of the paths, computation goes on and
does converge. In all topologies, once a critical threshold is exceeded,
convergence becomes impossible. This can be seen, in Figure \ref{fig:loss-util},
for the Barabasi-Albert topology at a loss rate of more than 1\% and
for the other topologies at a loss rate higher than 5\%. In further
experiments with a higher drop rate, Barabasi-Albert topology was
always the most sensitive to message loss and grid topology the least
sensitive. This supports the hypothesis on the effect of multiple
paths, because in grid topology every two neighbors are tightly connected
through their other neighbors.

\subsection{Data properties}

\begin{figure*}
\subfloat[\label{fig:bias-util}Sensitivity to bias-- convergence]{\includegraphics[width=0.5\textwidth]{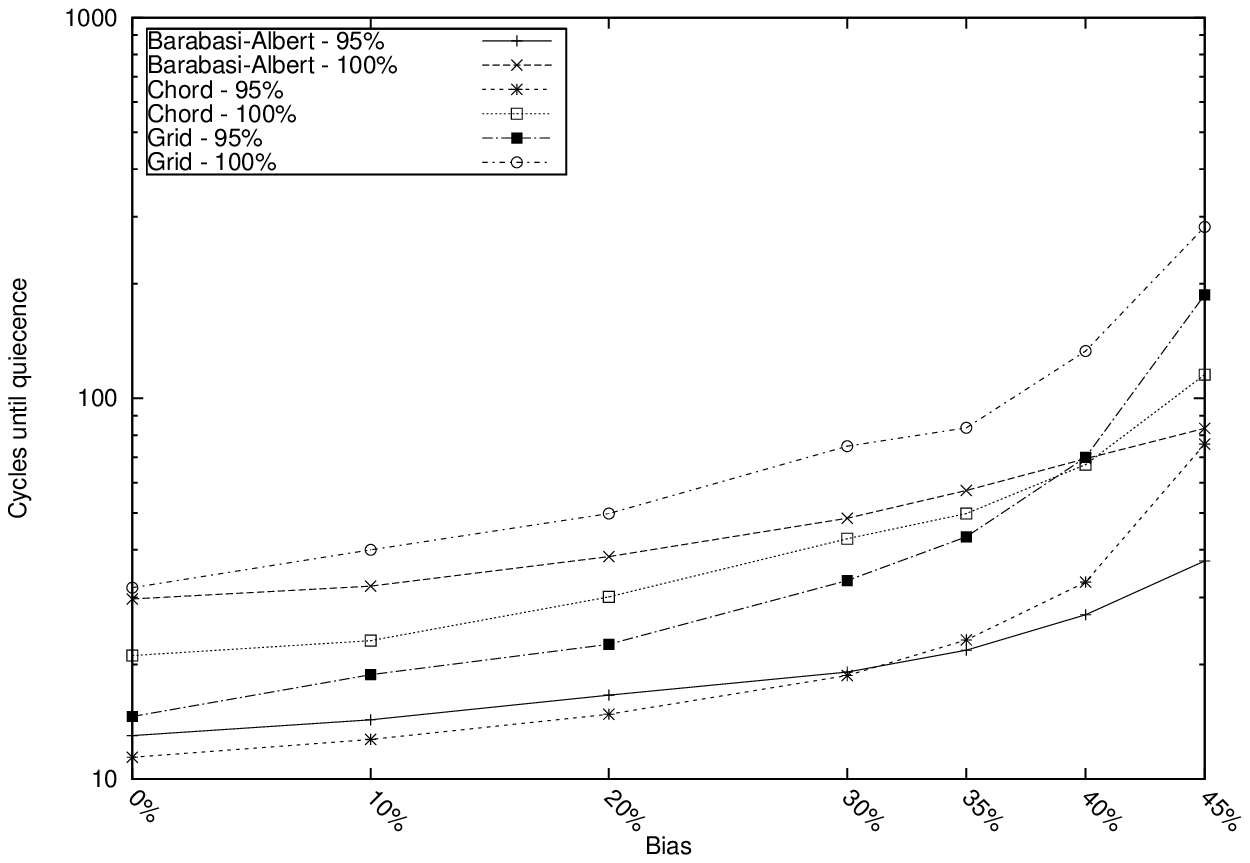}}\subfloat[\label{fig:bias-msgs}Sensitivity to bias -- communication]{\includegraphics[width=0.5\textwidth]{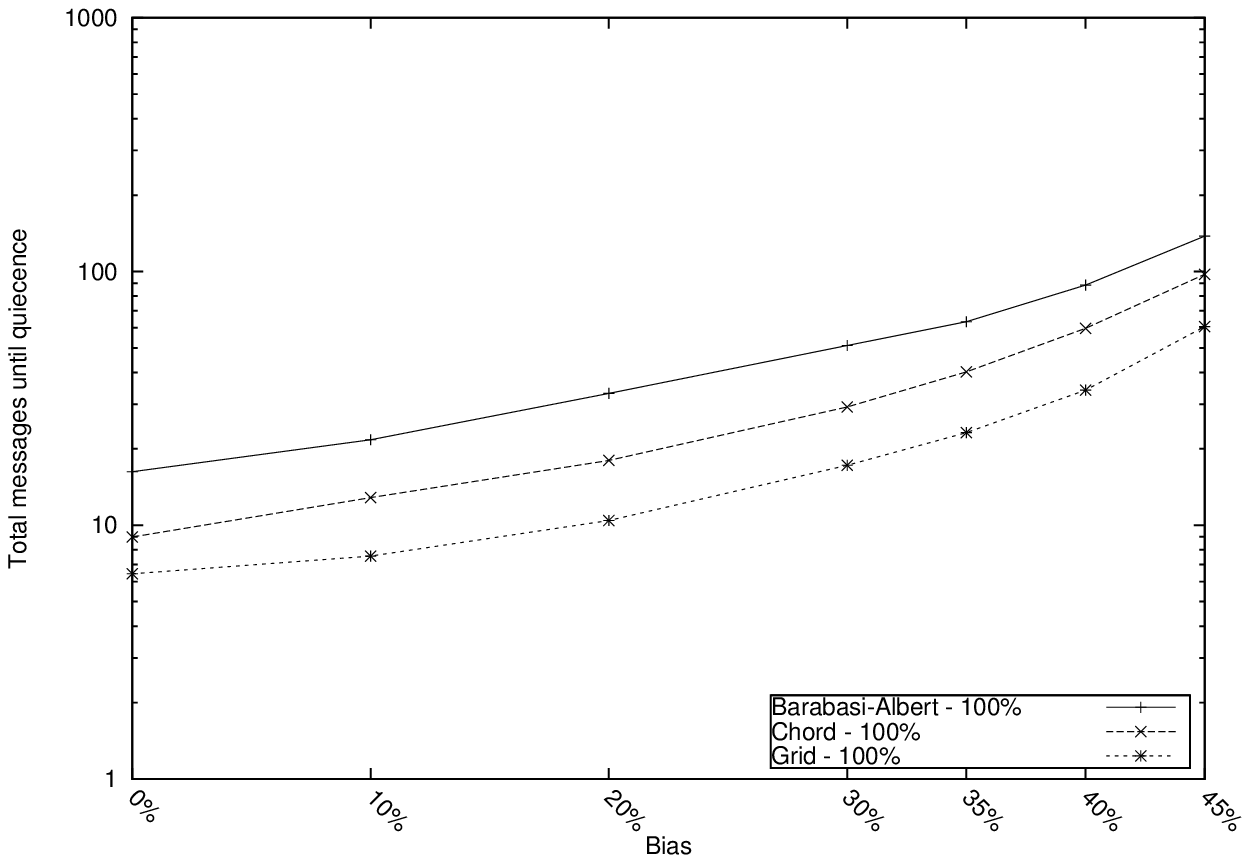}}

\subfloat[\label{fig:std-util}Sensitivity to variance -- convergence]{\includegraphics[width=0.5\textwidth]{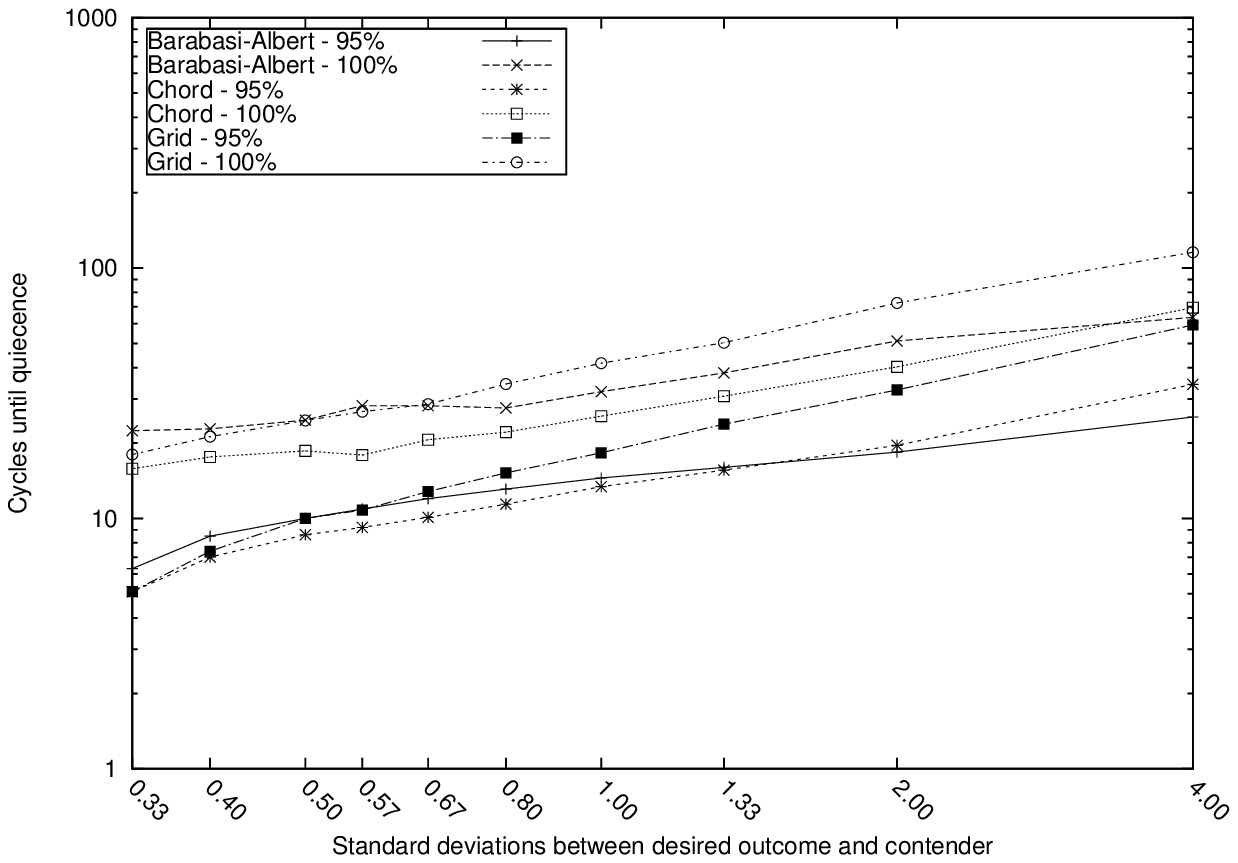}}\subfloat[\label{fig:std-msg}Sensitivity to variance -- communication]{\includegraphics[width=0.5\textwidth]{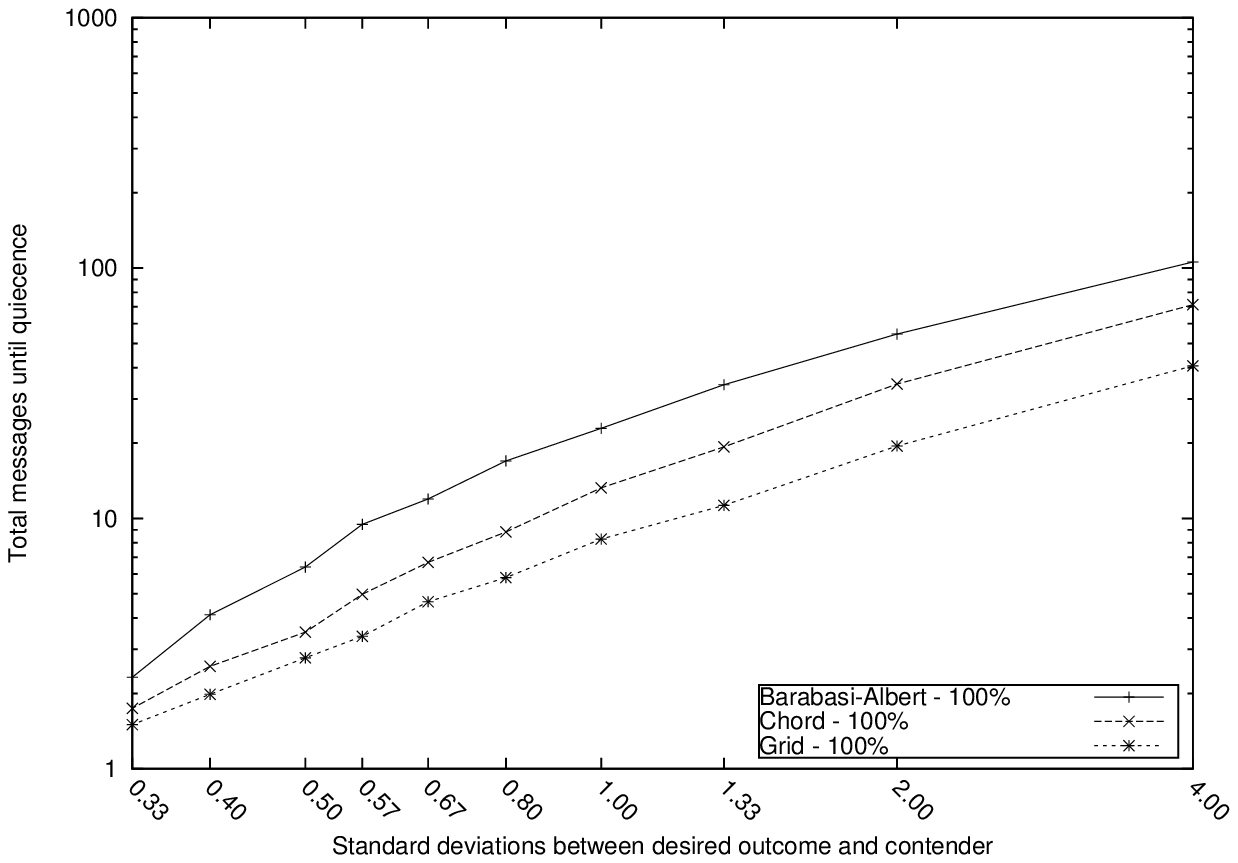}}

\caption{Problem difficulty}
\end{figure*}

Next, we investigate the sensitivity of LSS to the difficulty of the
problem. From previous research it is known that the performance of
local thresholding algorithms depends mainly on the proximity of the
average to the decision threshold. As can be seen in Figures \ref{fig:bias-util}
and \ref{fig:bias-msgs}, the communication overhead and the 95\%
convergence rate decrease super-exponentially with the bias. The 100\%
convergence rate also decreases with the bias, although perhaps not
exponentially.

Increased noise also makes computation more costly. Figures \ref{fig:std-util}
and \ref{fig:std-msg} show that as the standard deviation is increased
from a quarter of the distance between the desired outcome and the
contender to four times that distance, convergence time grows linearly
and the message overhead grows sublinearly.

\subsection{Ineffective parameters}

The number of possible solutions, $k$, and the dimensionality, $d$,
of the data, can also affect performance. However, experiments with
$k$ ranging from 3 to 243 revealed no sensitivity of the performance.
A similar result was obtained when the dimension of the data was varied
from $d=1$ through $d=6$. We conclude that, once bias and variance
are accounted for, neither the number of possible solutions nor the
dimensionality of the data has any bearing on performance. We chose
not to present these results graphically%
\footnote{As said earlier, code will be provided for anyone who wishes to verify
these results.%
}. 

The algorithm's parameters, the minimal weight allowed for $\left|S_{i}\right|$,
$\beta$, and the lower bound on the delay, $\ell$, were also experimented
with. In our experiments, the algorithm underperformed when both were
zero. However, setting the parameters to larger values than the default
had no noticeable effect. Thus, we keep $\beta=0.001$ and $\ell=1$,
and refrain from presenting numeric results.

\subsection{Dynamic data}

The next set of experiments is, arguably, the most realistic. In these
experiments the data at the peers was randomly changed at a controlled
rate for 100,000 simulation cycles. This hinders convergence and causes
a constant need for further communication. Thus, the average number
of peers which compute the wrong outcome and the average number of
messages per link per cycle are reported instead of the convergence
rate and total number of messages. Because of the large number of
events in these experiments, they were executed on networks of just
1000 peers. Also, these experiments were carried out with twice the
default bias (20\% rather than 10\%) and twice the default standard
deviation of the data (twice the distance of the desired outcome from
the contender rather than just once that distance) in order to increase
the effect of every change.

\begin{figure*}
\subfloat[\label{fig:dnoise-util-2}Average percent of errors]{\includegraphics[width=0.5\textwidth]{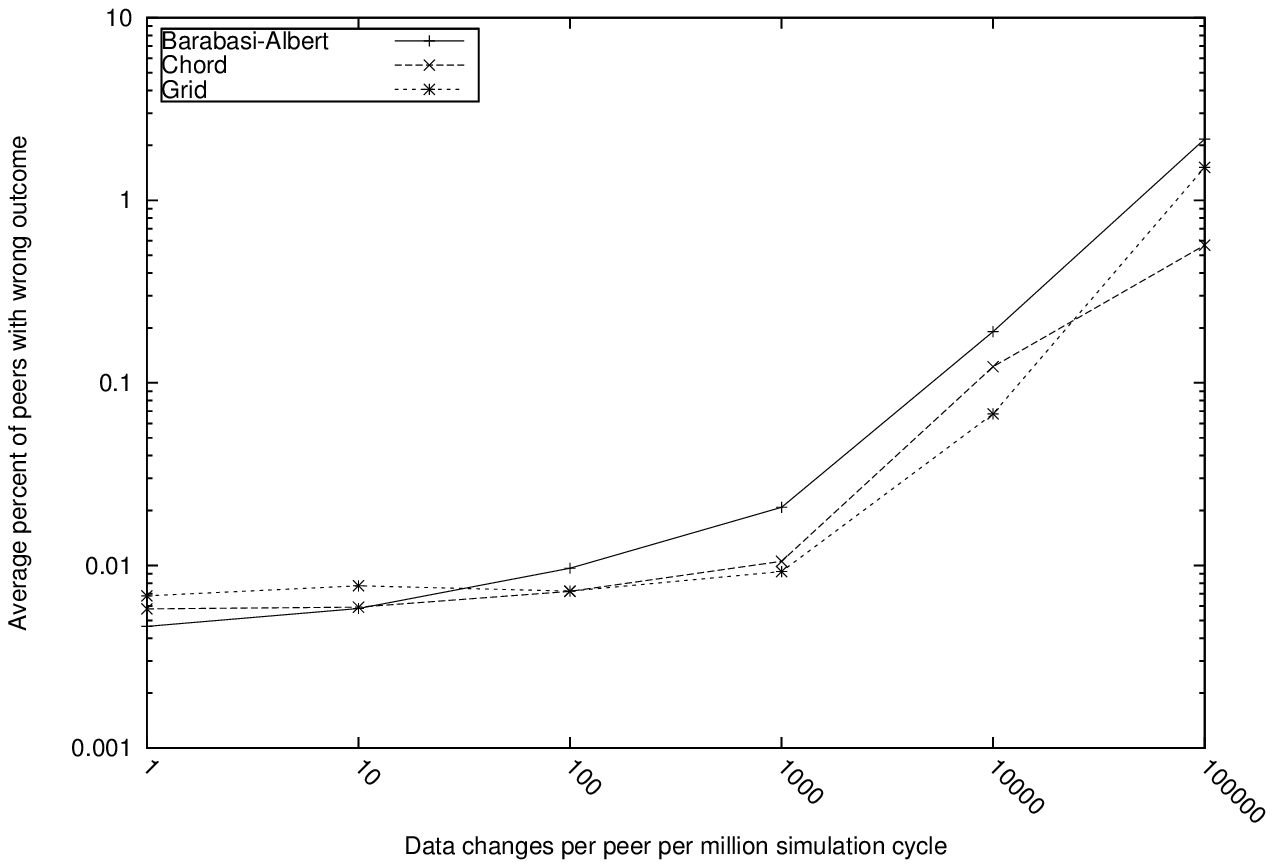}}\subfloat[\label{fig:dnoise-msgs-2}Average number of messages per cycle per
connection]{\includegraphics[width=0.5\textwidth]{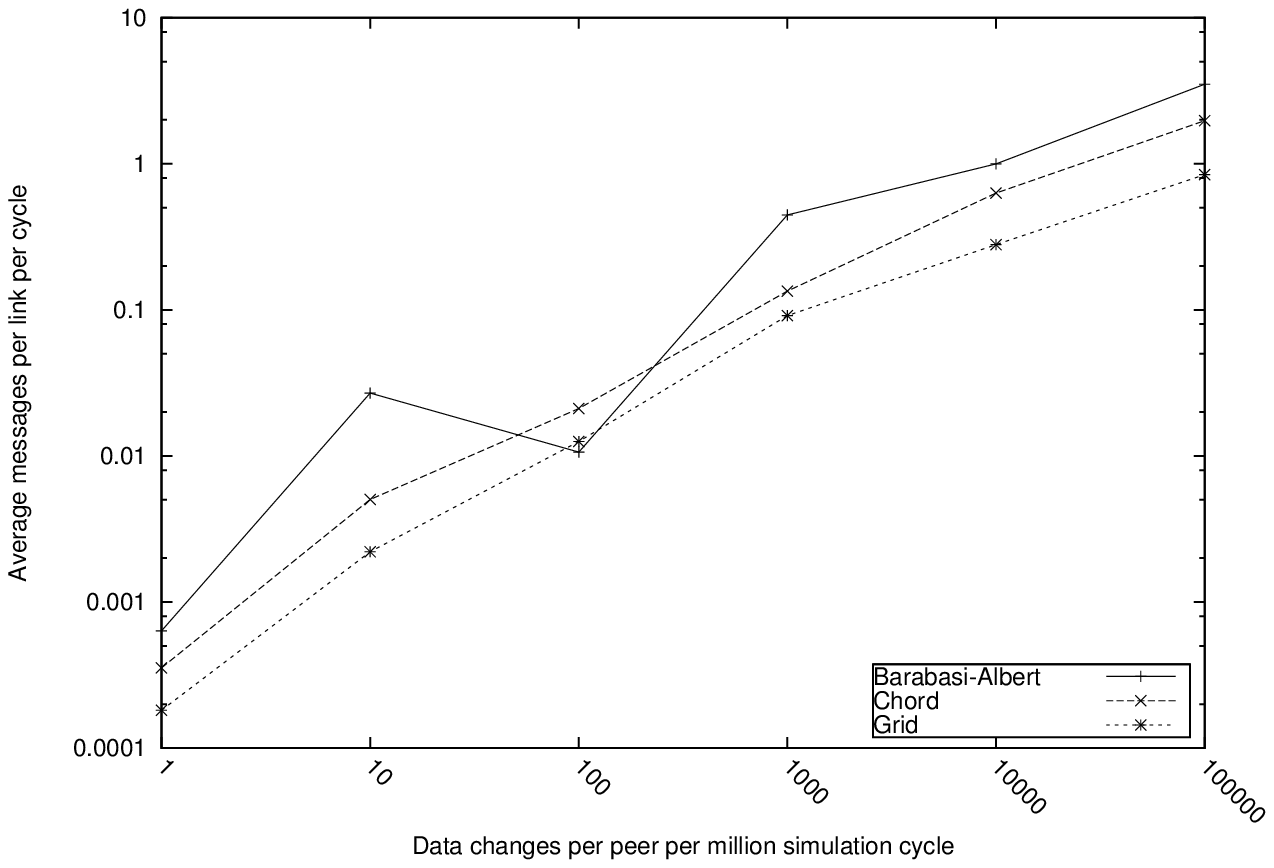}}\caption{Dynamically changing data}
\end{figure*}

As can be seen in Figures \ref{fig:dnoise-util-2} and \ref{fig:dnoise-msgs-2},
up to a noise rate in which one peer's input changes, on average,
at each simulator cycle, the effect of data dynamics is almost only
on communication and not on accuracy. Correction is, apparently, fast
enough so that the occasional changed input does not propagate an
error to a larger part of the network. On the other hand, that same
correction does cost in messages. So communication cost does grow
about linearly with the noise rate. Then, at about the noise rate
which provides for one change in each simulator cycle, the effect
of the change on accuracy begins to become noticeable, and the errors
accumulate linearly with the noise rate.

The effect of message loss in a dynamic setup is different than it
is when the data is static. In a static setup, a peer that does not
receive the intended message does not react to correct the wrong output
of the sender. In a dynamic setup, however, a peer has many more triggers
that will cause it to react, and message loss has only a short-term
effect on correctness.

\begin{figure*}
\subfloat[\label{fig:dloss-util}Average percent of errors]{\includegraphics[width=0.5\textwidth]{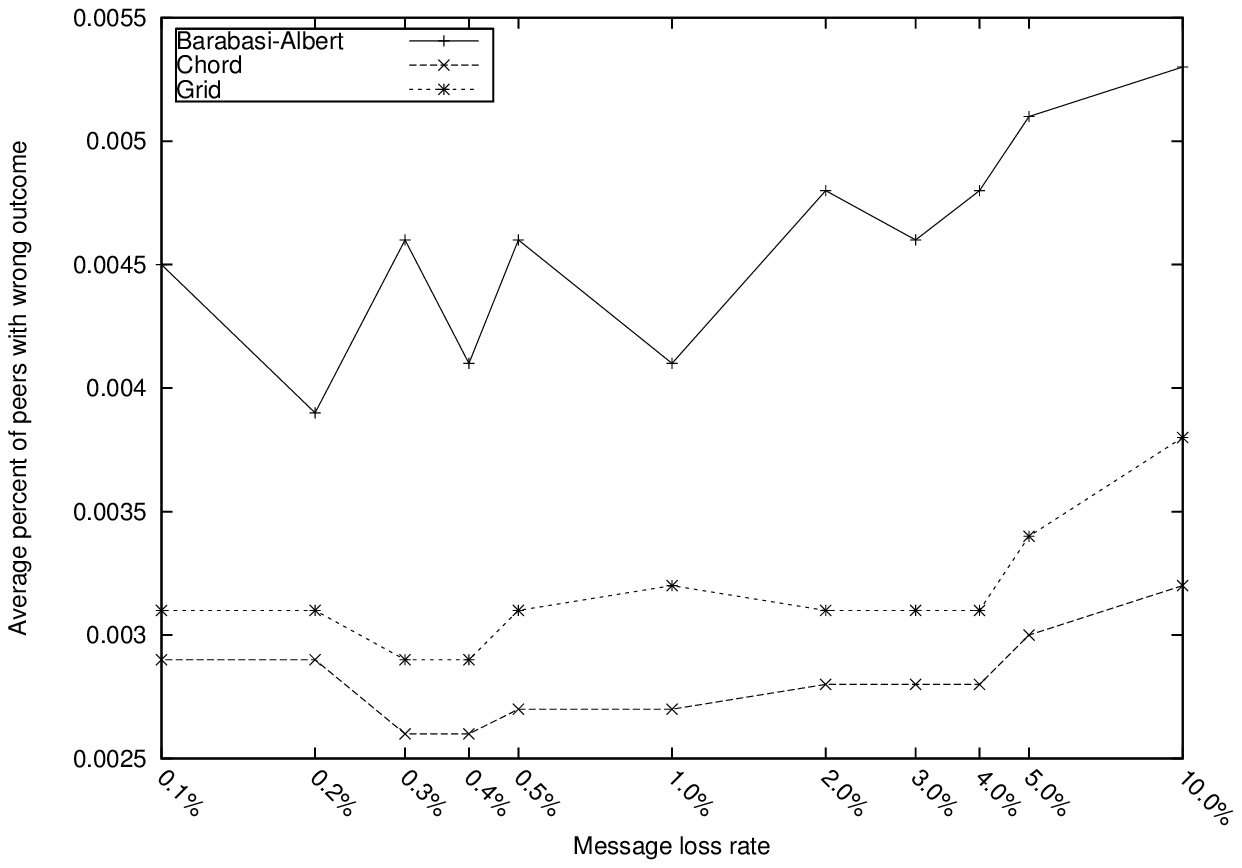}}\subfloat[\label{fig:dloss-msgs}Average number of messages per cycle per connection]{\includegraphics[width=0.5\textwidth]{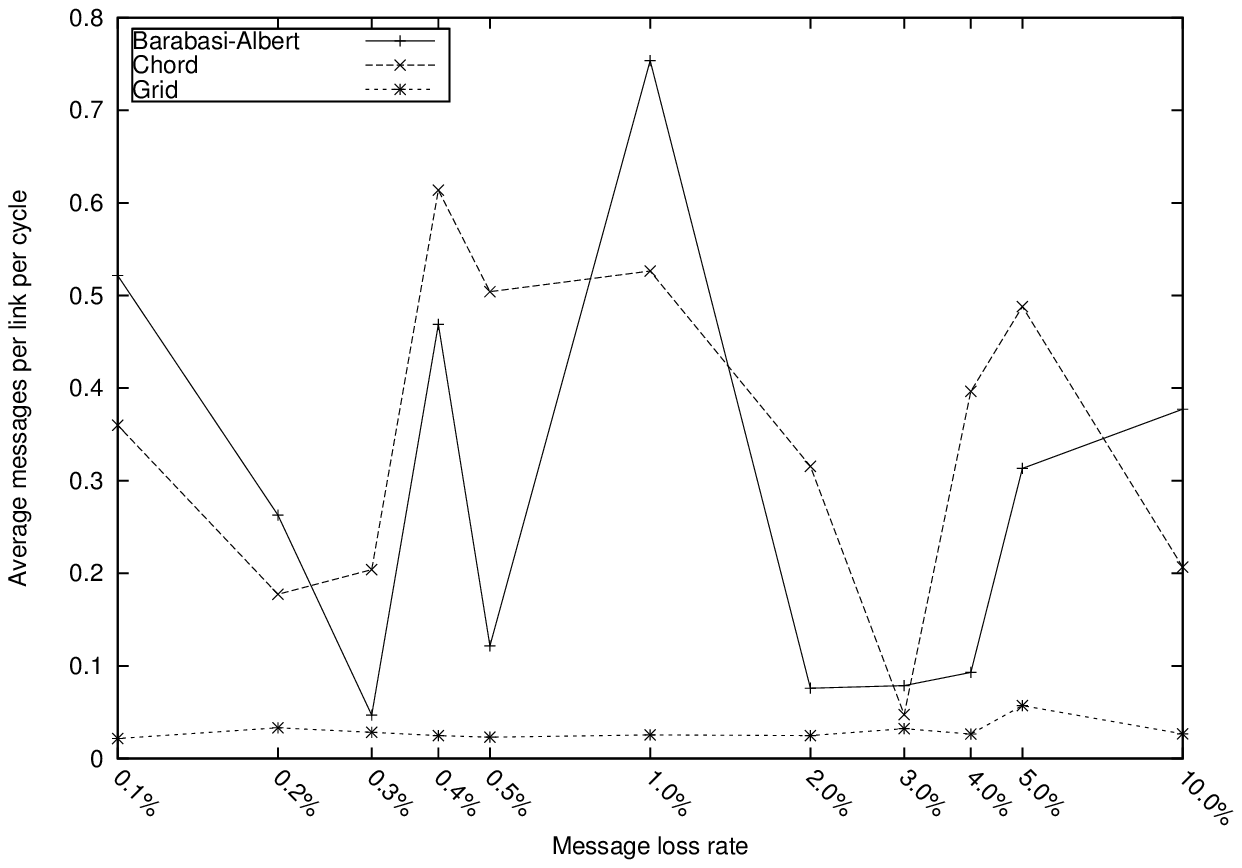}}\caption{Message loss and dynamically changing data}
\end{figure*}

Evidence for this can be seen in Figure \ref{fig:dloss-util} and
Figure \ref{fig:dloss-msgs}. In these experiments, the loss rate
is gradually increased, and the data is dynamically changed at a rate
of one thousand peers per million per simulator cycle. As can be seen,
the percentage of peers that compute the wrong result is extremely
low. This means that the errors induced by message loss hardly accumulate.
When 5\% of all messages are lost, the error rate is less than half
of a percent. In comparison, in Figure \ref{fig:loss-util}, one can
see that in an experiment in which the data is static, the error rate
skyrockets at this loss rate.

The second phenomenon evident in this dynamic setup is that the variance
of both the accuracy (Figure \ref{fig:dloss-util}) and the communication
overhead (Figure \ref{fig:dloss-msgs}) is very large in Barabasi-Albert
and in Chord topologies, but not in the Grid topology. Again, our
hypothesis is that a greater number of short alternate routes between
every two neighbors increases the algorithm's robustness to message
loss. 

\subsection{Dynamic network}

Finally, the robustness of the LSS algorithm to peer churn is validated.
Again, a network of 2,000 peers is simulated for 100,000 cycles with
their data changing at a rate of 1,000 ppmc. Additionally, peers drop
out of the system at a controlled rate of between zero (no churn)
and four ppmc. It is assumed that churn is detected by the peer's
neighbors, which then recalculate their status and correct as needed.

\begin{figure*}
\subfloat[\label{fig:churn-util}Average percent of errors]{\includegraphics[width=0.5\textwidth]{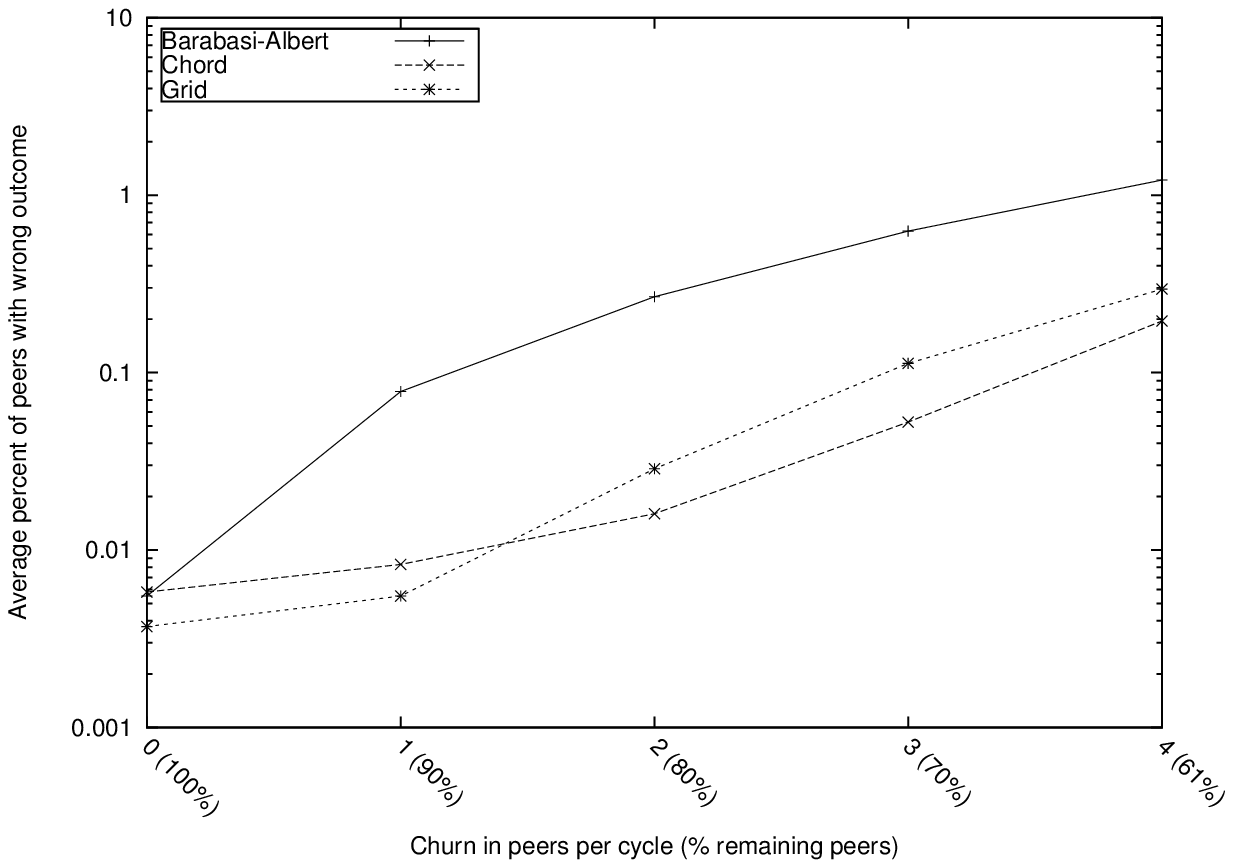}}\subfloat[\label{fig:churn-msgs}Average number of messages per cycle per connection]{\includegraphics[width=0.5\textwidth]{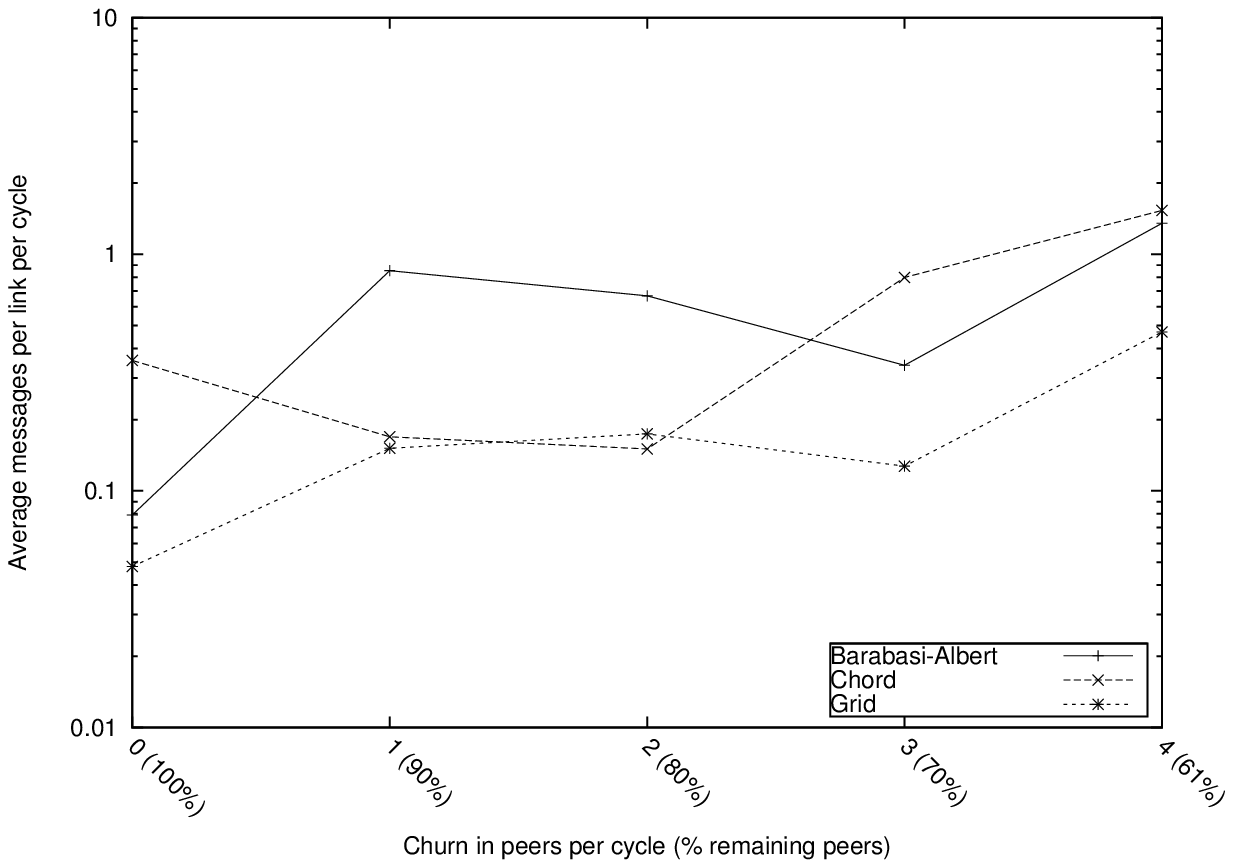}}\caption{Churn and dynamically changing data}
\end{figure*}

Figures \ref{fig:churn-util} and \ref{fig:churn-msgs} depict the
effect different churn rates have on the average error and the average
message load per link. Beside the churn rate, the x-axis also denotes
the percentage of peers remaining active at the end of the 100,000
cycles. As can be seen in Figure \ref{fig:churn-util}, the error
rate grows notably as more peers churn. However, even when eventual
churn nears 40\% of the peers, no more than 1\% of them compute the
wrong outcome on average. Message overhead increases with churn, probably
due to the increased effort needed to correct the mistaken outcome.
However, the trend is not very clear and the overhead is very noisy.
It is worth noting that the regularity of the bi-dimensional grid
is degraded with churn, which may explain why performance is similar
in that topology and in the other two topologies even though they
are less regular.

\section{\label{sec:Related-work}Related work}

This paper makes a fundamental contribution to computation in large
distributed systems. As such, it relates to other methods of computation
in similar networks. We choose to categorize such methods according
to the regime used to dictate which messages are sent.

The first category is algorithms that enforce a strict messaging regime.
This category includes convergecast-based in-network computation (e.g.,
those using MapReduce \cite{disco}), which have been used for years
in small-scale distributed systems. It also includes algorithms such
as \cite{Unnamed-3}, in which messages flow through the entire system
in a strict order. As systems grow larger, global methods lose their
appeal. This is mainly because enforcing order, synchronization, and
reliability becomes impermissibly costly.

The second category is algorithms that are based on repeated averaging.
General results about the convergence of statistics under repeated
averaging are known since the 70's \cite{Unnamed-2}. They were first
implemented for function computation in a distributed system in applications
such as diffusive load balancing \cite{diffusiveLB}, averaging \cite{averagingIterations},
and Kalman filtering \cite{gossipKalman0}. The first relation of
repeated averaging to distributed data mining in peer-to-peer system
was apparently in the context of the DREAM project \cite{newscast}.
Diffusion has also been shown to allow solving more general optimization
problems than merely averaging \cite{subgradient,chen12}.

Kempe et al., however, were the first to position repeated averaging
in the context of gossip algorithms and to provide much needed bounds
on convergence speed \cite{KempeGossip,gossipBoyd,Gossip,GossipVoting,dynamicGossip}.
Gossip based algorithms were shown to converge with the logarithm
of the network size if uniformly random messaging is possible. Otherwise,
their convergence rate depends on the eigen-gap of the network graph
\cite{eigengap}. Gossip-based algorithms are also simple and robust
and have been applied to a large number of problems (see, e.g., \cite{gossipEM,gossipSL}).
However, since, at base, their convergence rate depends on the random
mixing of inputs, gossip algorithms are still extremely wasteful.
In a wireless sensor network, where the messaging budget is scarce,
it seems inherently wrong to send messages at random.

Local thresholding defines the last category of algorithms. Unlike
gossip-based algorithms, local algorithms are deterministic. They
are also far more data dependent. There are ample situations in which
a peer running local thresholding will not send any message at all.
Comparative testing of local thresholding and gossip algorithms \cite{726272}
shows that the former are vastly more efficient. 

Previous local thresholding algorithms all rely on the cycle freedom
of the network. The proof of the stopping condition they use relies
on the fact that the latest message received from a neighbor $p_{j}\in N_{i}$
does not depend on the input of $p_{i}$ or on inputs which are accounted
for in messages received from other $p_{k}\in N_{i}$. 

Besides the difficulty of providing a cycle free network, this also
meant those algorithms were critically dependent on the reliability
of messages to neighbors. If a message is lost on the way from $p_{i}$
to $p_{j}$ then there is no alternative path in which the inputs
represented in that message can propagate the $p_{j}$. This dependency
has made them even less suitable to wireless sensor network applications.
This work is the first to lift the requirement of cycle freedom.

\section{\label{sec:Conclusion-and-further}Conclusion and further research}

The new local stopping rule and update methods presented in this paper
remove a difficult barrier to the implementation of local thresholding
algorithms in some of the most popular distributed environments. It
is further hoped that the novel presentation of these algorithms as
operating in the field of weighted vectors will simplify future developments.
As demonstrated here, local thresholding can be an extremely efficient
way to compute complex functions over large distributed networks,
regardless of their topological characteristics.

Recent years have seen a lot of focus on applications of local thresholding
algorithms and on their stopping rules. This work, and especially
the presentation of the problem in terms of weighted vectors, greatly
simplifies the argument in favor of certain update policies. We believe
some interesting problems lie in the correction policy. Although much
important work has been done on expediting convergence of gossip algorithms
and diffusive load balancing \cite{diffusiveLB} algorithms using
smart update rules, we are not aware of any parallel work on local
thresholding algorithms.

The generalization described here still misses one interesting aspect
of some of the target systems: that the communication graph is often
directed and weighted. In a wireless sensor network, the signal transmitted
by a sensor may well be received by a sensor whose signal it cannot
receive. The might also have different energy levels, meaning messages
are more costly to one than they are to the other. In structured peer-to-peer
systems such as Chord, routing from a peer to a peer in its finger
table costs just one message whereas routing in the opposite direction
can cost $\log\left(n\right)$ messages. In general, expediting convergence
by taking into account message delays, and not merely connectivity,
may be an important challenge.

\section{Acknowledgement}

This research was supported, in part, by the Network Science and Technology
Center at Rensselaer Polytechnic Institute. The author wishes to thank
the people at NeST for their hospitality.

\begin{figure}[H]
\includegraphics[width=0.2\textwidth]{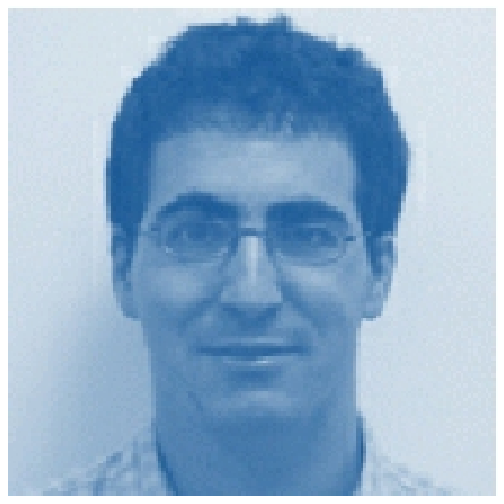}Ran Wolff is a member of
the information systems department at university of Haifa, Israel.
His main fields of research are large-scale data mining, both widely
distributed or in fast streams, and privacy preserving data mining.
Ran develops algorithms suitable for peer-to-peer networks, grid systems,
and wireless sensor networks, and contributes to the core definitions
of the meaning of privacy in a data intensive environment. Ran regularly
serves as PC in the ICDM, SIGKDD, and WWW conferences, and as a reviewer
for the DMKD and TKDE journals, among other.

\end{figure}

\end{document}